\documentclass[twocolumn]{emulateapj}

\lefthead{Law \& Majewski 2010}
\righthead{Galactic Satellites of Sgr}

\slugcomment{DRAFT: \today}

\begin{document}

\newcommand{\msun}{\ensuremath{\rm M_\odot}}
\newcommand{\msunyr}{\ensuremath{\rm M_{\odot}\;{\rm yr}^{-1}}}
\newcommand{\Ha}{\ensuremath{\rm H\alpha}}
\newcommand{\Hb}{\ensuremath{\rm H\beta}}
\newcommand{\lya}{\ensuremath{\rm Ly\alpha}}
\newcommand{\Ntwo}{[\ion{N}{2}]}
\newcommand{\kms}{\textrm{km~s}\ensuremath{^{-1}\,}}
\newcommand{\ztwo}{\ensuremath{z\sim2}}
\newcommand{\zthree}{\ensuremath{z\sim3}}
\newcommand{\feh}{\textrm{[Fe/H]}}
\newcommand{\afeh}{\textrm{[$\alpha$/Fe]}}
\newcommand{\nifeh}{\textrm{[Ni/Fe]}}

\newcommand{\hst}{{\it HST}-ACS}

\shortauthors{Law \& Majewski}
\shorttitle{Galactic Satellites of Sgr}

\title{Assessing the Milky Way Satellites Associated with the Sagittarius Dwarf Spheroidal Galaxy}

\author{David R.~Law\altaffilmark{1,2}, Steven R. Majewski\altaffilmark{3}}

\altaffiltext{1}{Hubble Fellow.}
\altaffiltext{2}{Department of Physics and Astronomy, University of California, Los Angeles, CA 90095;
drlaw@astro.ucla.edu}
\altaffiltext{3}{Dept. of Astronomy, University of Virginia,
Charlottesville, VA 22904-4325 (srm4n@virginia.edu)}

\begin{abstract}

Numerical models of the tidal disruption of the Sagittarius (Sgr) dwarf galaxy have recently
been developed that for the first time simultaneously satisfy most observational constraints on the angular position, distance,
and radial velocity trends of both leading and trailing tidal streams emanating from the dwarf.
We use these dynamical models in combination with extant 3-D position and velocity data for Galactic globular clusters and dSph galaxies
to identify
those Milky Way satellites that are likely to have originally formed in the gravitational potential
well of the Sgr dwarf, and have been stripped from Sgr during its extended interaction with the Milky Way.
We conclude that the globular clusters Arp 2,
M 54, NGC 5634, Terzan 8, and Whiting 1 are almost certainly associated with the Sgr dwarf,
and that Berkeley 29, NGC 5053, Pal 12, and Terzan 7 are likely to be as well (albeit at lower confidence).  
The initial Sgr system therefore
may have contained 5-9 globular clusters, 
corresponding to a specific frequency $S_N = 5 - 9$ for an initial Sgr luminosity $M_V = -15.0$.
Our result is consistent with the $8\pm2$ genuine Sgr globular clusters expected on the basis of statistical modeling of the Galactic globular cluster distribution
and the corresponding false-association rate due to chance alignments with the Sgr streams.
The globular clusters identified as most likely to be associated with Sgr are consistent with previous reconstructions of the Sgr age-metallicity relation,
and show no evidence for a second-parameter effect shaping their horizontal branch morphologies.
We find no statistically significant evidence to suggest that any of the recently discovered population of ultra-faint dwarf galaxies are associated with the Sgr tidal streams,
but are unable to rule out this possibility conclusively for all systems.

\end{abstract}


\section{INTRODUCTION}

Globular clusters tend to form in all sufficiently massive galactic systems during episodes of major star formation, their relatively simple stellar populations providing important constraints
on starburst and galaxy formation models (see, e.g., Brodie \& Strader 2006).
While no globular clusters are found in low mass ($M_V \gtrsim -12$) dwarf galaxies,
the more massive dwarfs
such as the LMC ($M_V = -18.5$),
SMC ($M_V = -17.1$) and Fornax ($M_V = -13.1$) are observed to contain
nineteen, eight, and five  globular clusters respectively (see, e.g., Forbes et al. 2000).
It is generally accepted that
the Sagittarius (Sgr) dwarf spheroidal galaxy ($M_V = -13.64$; Law \& Majewski 2010 [hereafter LM10]) contains at least four clusters
(M 54, Arp 2, Terzan 7, Terzan 8; Da Costa \& Armandroff 1995) within its main body.
Since Sgr is known to have experienced significant tidal mass loss, however, it is likely that additional globular clusters 
may have been stripped from Sgr during its prolonged interaction with the Milky Way and now lie scattered throughout
the Galactic halo.

In the anticipation that such clusters are likely distributed across the sky, various authors
(e.g., Lynden-Bell \& Lynden-Bell 1995; Irwin 1999; Bellazzini et al. 2002; Newberg et al. 2003; Majewski et al. 2004; Carraro et al. 2007) 
have suggested a number of additional globular clusters 
(e.g., Pal 2, Pal 12, NGC 5634, NGC 2419, Whiting 1, etc.) that may have formed in the Sgr dSph (see Table \ref{prevcands.table} for a summary).  
Perhaps the most comprehensive efforts to conduct a systematic census to date have been undertaken by Palma et al. (2002) and Bellazzini et al. (2003a),
who demonstrated that there is a strong correlation between the positions and radial velocities of outer halo (Galactocentric radius $r_{\rm GC} > 10$ kpc) 
globular clusters and the orbit of Sgr.
Bellazzini et al. (2003a) in particular presented a statistical assessment of the significance of the total number of Galactic globular clusters aligned with the Sgr orbital plane,
concluding that there was a  $\lesssim 2$\% probability that the observed grouping of clusters about the Sgr plane arises by chance.
While this approach was certainly effective, it was based on the projected {\it orbit} of Sgr (as derived by Ibata \& Lewis 1998) and therefore 
did not adopt an optimal membership criterion because
tidal debris does not precisely follow the orbit of the parent object:
Trailing tidal arms lie outside the orbital path of the parent while
leading tidal arms lie interior to the orbit and, in the case of Sgr,
wrap up around the Galactic Center (see discussion by Johnston et al. 1995, 1999; Choi et al. 2007; Eyre \& Binney 2009).
Moreover, Bellazini et al. (2003a) concentrated on the past orbit of Sgr, an approach that
misses potential connections to the leading arm tidal debris.

In this contribution we wish to update and extend these previous analyses,
and ask whether each individual Milky Way satellite matches the
angular position, distance, and radial velocity of Sgr tidal debris
sufficiently well that it is statistically likely to be physically associated with the stream and might therefore have originated in the Sgr dwarf.
Such an undertaking is greatly aided by the significantly improved picture of the Milky Way --- Sgr system that has been provided in recent years
by the Two Micron All-Sky Survey (2MASS) and Sloan Digital Sky Survey (SDSS) which have 
mapped the stellar streams from the Sgr dwarf wrapping fully $360^{\circ}$ across the sky
(Majewski et al. 2003; Belokurov et al. 2006a; Yanny et al. 2009).  These surveys have also revealed a wealth of additional substructure in the Galactic halo
in the form of both globular clusters and a population of dwarf galaxies with masses $\sim 10^7 M_{\odot}$ within their central 300 pc (Strigari et al. 2008)
but ultra-faint luminosities comparable to those of globular clusters.
The origin of these ultra-faint dwarfs is unknown, and
we consider whether any of them may be dynamically associated with the Sgr dSph.

Our effort is also aided by the recent construction of a 
numerical model of the Sgr stream that
reproduces the majority of the observed characteristics of the stellar tidal streams with extremely high fidelity (Law et al. 2009; LM10).
The comprehensive view of both leading and trailing arm material (which overlap each other
along the line of sight for large swathes of the sky) provided by this model allows us
to view potential Sgr members in the context of the full debris system, and assess possible associations with leading and trailing arms based not only on spatial distributions, but 
on dynamical properties such as radial velocities and proper motions as well.

\begin{deluxetable*}{llccccccccc}
\tablecolumns{11}
\tablewidth{0pc}
\tabletypesize{\scriptsize}
\tablecaption{Candidate Sgr Stream Globular Clusters Proposed in the Literature\tablenotemark{a}}
\tablehead{\colhead{ID} & \colhead{Name} & \colhead{DA95} & \colhead{LL95} & \colhead{I99}
& \colhead{D01} & \colhead{P02} & \colhead{B03} 
& \colhead{M04} & \colhead{C07} & \colhead{LM10b}}
\startdata
45	&	Arp 2		& YES	& YES	& YES	& ...		& YES	& YES	& YES	& ...		& YES\\%
52	&	M 2			& ...		& ...		& ...		& ...		& ...		& YES	& ...		& ...		& NO\\
...	&	M 5			& ...		& ...		& ...		& ...		& YES	& ...		& ...		& ...		& NO\\
21	&	M 53			& ...		& ...		& ...		& ...		& YES	& ...		& ...		& ...		& NO\\
42	&	M 54			& YES	& YES	& YES	& ...		& YES	& YES	& YES	& ...		& YES\\%
1	&	NGC 288		& ...		& ...		& ...		& ...		& ...		& YES	& ...		& ...		& NO\\
13	&	NGC 2419	& ...		& ...		& YES	& ...		& ...		& ...		& ...		& ...		& NO\\
18	&	NGC 4147	& ...		& ...		& ...		& ...		& NO	& YES	& ...		& ...		& NO\tablenotemark{b}\\%
22	&	NGC 5053	& ...		& ...		& ...		& ...		& NO	& YES	& ...		& ...		& YES\\%
26	&	NGC 5466	& ...		& ...		& ...		& NO	& YES	& YES	& ...		& ...		& NO\\
27	&	NGC 5634	& ...		& ...		& ...		& ...		& ...		& YES	& ...		& ...		& YES\\%
30	&	NGC 5824	& ...		& ...		& ...		& ...		& ...		& YES	& ...		& ...		& NO\\
40	&	NGC 6426	& ...		& ...		& ...		& ...		& ...		& YES	& ...		& ...		& NO\\
...	&	NGC 6356	& ...		& ...		& ...		& ...		& YES	& ...		& ...		& ...		& NO\\
7	&	Pal 2			& ...		& YES	& YES	& ...		& ...		& YES	& YES	& ...		& NO\tablenotemark{b}\\%
31	&	Pal 5			& ...		& ...		& ...		& ...		& NO	& YES	& ...		& ...		& NO\\
53	&	Pal 12		& ...		& ...		& YES	& ...		& YES	& YES	& YES	& ...		& YES\\%
54	&	Pal 13		& ...		& ...		& ...		& ...		& NO	& YES	& ...		& ...		& NO\\
19	&	Rup 106		& ...		& ...		& ...		& ...		& ...		& YES	& ...		& ...		& NO\\
...	&	Terzan 3		& ...		& ...		& ...		& ...		& ...		& YES	& ...		& ...		& NO\\
44	&	Terzan 7		& YES	& YES	& YES	& ...		& YES	& YES	& YES	& ...		& YES\\%
46	&	Terzan 8		& YES	& NO	& YES	& ...		& YES	& YES	& YES	& ...		& YES\\%
3	&	Whiting 1		& ...		& ...		& ...		& ...		& ...		& ...		& ...		& YES 	& YES\\%
\enddata
\tablenotetext{a}{Columns represent Da Costa \& Armandroff (1995; DA95), Lynden-Bell \& Lynden-Bell (1995; L95), Irwin et al. (1999; I99), 
Dinescu et al. (2001; D01), Palma et al. (2002; P02),
Bellazzini et al. (2003a; B03), Majewski et al. (2004; M04), Carraro et al. (2007; C07), this contribution (LM10b).  `YES' indicates possible candidate, `NO' indicates disfavored
by indicated study.}
\tablenotetext{b}{As discussed in \S \ref{discussion.sec}, NGC 4147 and Pal 2 may plausibly be associated with the Sgr stream, but at low statistical confidence.}
\label{prevcands.table}
\end{deluxetable*}

This paper is organized as follows.  In \S \ref{model.sec} we describe the general characteristics of the Sgr tidal streams and the $N$-body model that best reproduces its observed
properties.  In \S \ref{gcsel.sec} we describe our
sample of Galactic satellites, incorporating globular clusters, ultra-faint dwarfs, and two open clusters for which association with Sgr has previously been claimed in the literature.
In \S \ref{quantifying.sec} we describe our numerical technique for quantifying the association of a Galactic satellite with the Sgr stream, applying this to individual
stellar clusters and ultra-faint dwarfs in \S \ref{compareGC.sec} and \S \ref{compareUF.sec} respectively.
We discuss the resulting implications for the star cluster budget of Sgr and its contributions to the Galactic halo in \S \ref{discussion.sec}, summarizing our
conclusions in \S \ref{summary.sec}.

We adopt the heliocentric Sgr coordinate system ($\Lambda_{\odot}, B_{\odot}$) defined by Majewski et al. (2003) in which the longitudinal coordinate $\Lambda_{\odot} = 0^{\circ}$
in the direction of Sgr and increases along trailing tidal debris, and $B_{\odot}$ is positive towards the orbital pole $(l,b)_{\rm pole} \approx (274^{\circ}, -14^{\circ})$ (see
discussion by LM10).  All radial velocities are given in the Galactic Standard of Rest (GSR) frame, with respect to which the Sun has a peculiar motion
($U,V,W$) = ($9,12+220,7$) km s$^{-1}$ and is located $R_{\odot} = 8$ kpc from the Galactic Center.

\section{THE SAGITTARIUS DISRUPTION MODEL}
\label{model.sec}

We elect to compare the Galactic satellite population to the LM10 $N$-body model\footnote{This model is available online at http://www.astro.virginia.edu/$\sim$srm4n/Sgr/} 
of the Sgr streams.  As discussed at length in LM10, this model was constrained to match the
abundant observational data  on the angular positions (e.g., Majewski et al. 2003; Belokurov et al. 2006a; Yanny et al. 2009), radial velocities (e.g., Law et al. 2004, 2005;
Majewski et al. 2004; Monaco et al. 2007; Yanny et al. 2009), and chemical abundances (Chou et al. 2007, 2010; Monaco et al. 2007; Yanny et al. 2009) 
of stars in the Sgr tidal streams which have recently been provided by 2MASS and SDSS.
This model was fully constrained using only angular positions and radial velocities for Sgr stream stars; photometric parallaxes were not 
used as a constraint because of their large observational uncertainties and systematic effects due to
variations in metallicity and age along the tidal arms (e.g., Chou et al. 2007, 2010).
As detailed in LM10, however, the $N$-body model matches distance estimates for stars in the Sgr streams to within observational uncertainty.


We fit to the LM10 $N$-body model instead of the raw observational data for four reasons:
(1) The positional data for the Sgr M-giant stream (Majewski et al. 2003) contains an intrinsic distance
scatter (estimated to be 17\% by Majewski et al. 2003) from the intrinsic width of the color-magnitude relation,
as well as poorly known systematic effects in distance because of variations in the metallicity distribution function of Sgr stream stars.
Similar systematic uncertainties also affect the SDSS observations, making
certain sections of the tidal arms difficult to convert to distance directly.  The LM10 model
incorporates these metallicity effects and is fully consistent with the M-giant color-magnitude relation and apparent magnitude trend.
(2) The available velocity data are still
incomplete, with whole sections of the tidal arms poorly sampled.  The LM10 model can fill in these gaps in our knowledge.
(3) There is a limit to the length of the Sgr tidal tails that
can be traced with individual stellar populations such as M-giants (Majewski et al. 2003) or blue horizontal branch stars (Yanny et al. 2009).
Additional tidal debris from older stellar generations may exist and simply be poorly traced by current observational data.
With the LM10 model, the observed tails may be extrapolated to include such hypothetical older tidal debris.
(4) The LM10 model can provide proper motion information at all positions along the Sgr tidal streams.  These proper motions are well-constrained by observations in the other four dimensions
of phase space, but are extremely difficulty to measure directly.

\begin{figure*}
\plotone{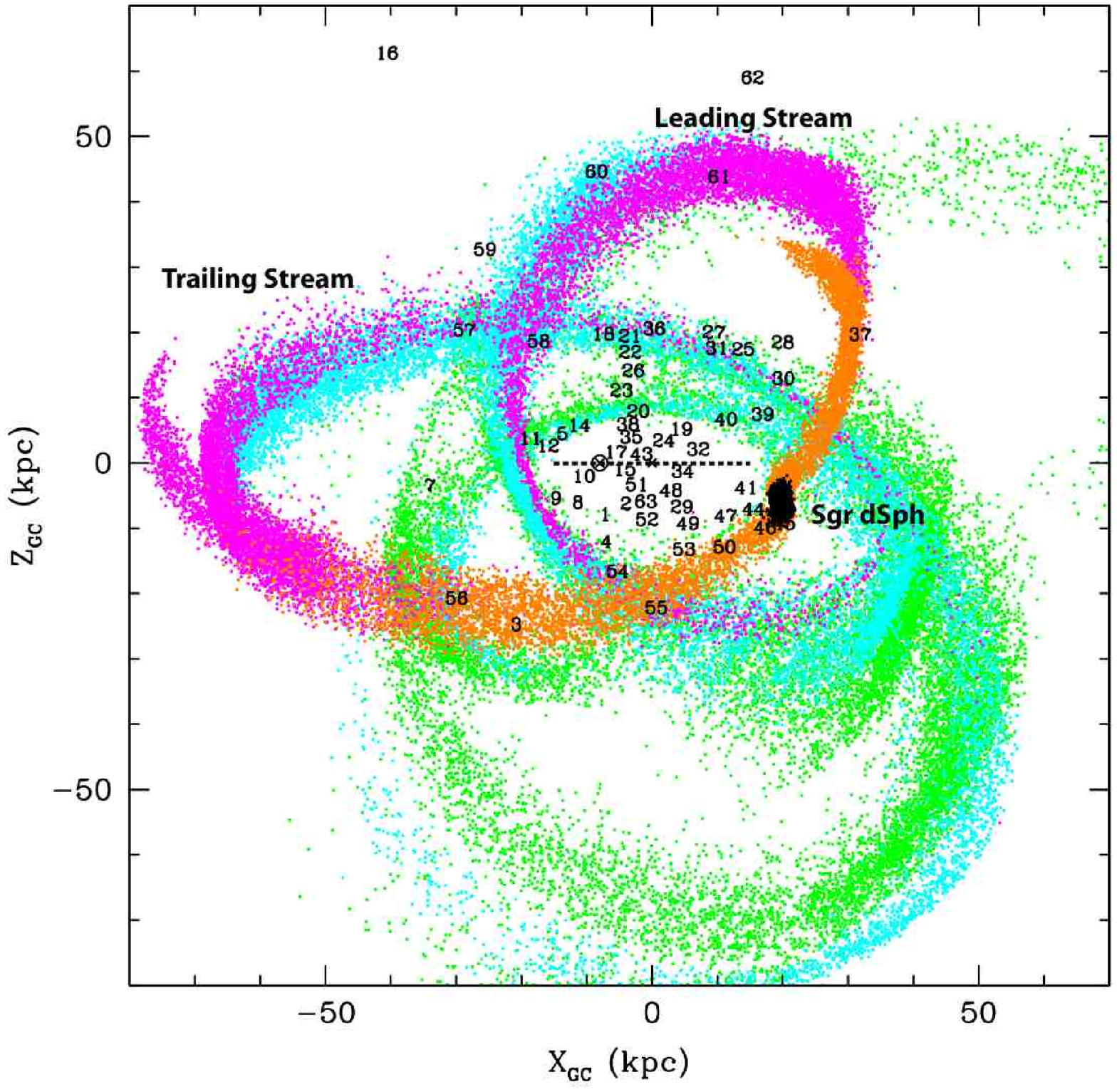}
\caption{Diagram showing the LM10 model of the Sgr system (colored points) in Galactocentric Cartesian coordinates ($X_{\rm GC}, Z_{\rm GC}$).  The color of individual
debris particles represents the time at which the particle became unbound from the Sgr dwarf.  The $\times$ indicates the location of the Galactic Center, while the circled $\times$
indicates the location of the Sun.  The horizontal dashed line represents the Galactic disk plane, the length of the line is chosen to indicate
a diameter 30 kpc.  Overplotted against the model are ID numbers corresponding to 
the locations in the $X_{\rm GC}$ --- $Z_{\rm GC}$ plane of Milky Way stellar satellites (ID numbers are matched with the corresponding satellite names in Table \ref{gcprops.table}).
The locations of some ID numbers have been adjusted slightly to minimize confusion from overlapping labels.}
\label{XYfig.fig}
\end{figure*}

In Figure \ref{XYfig.fig} (colored points) we plot a cross-sectional view of the LM10 model of the Sgr system in Galactocentric Cartesian coordinates.
The model satellite has been orbiting in the Galactic potential for $\sim 8$ Gyr, and has produced multiple epochs of tidal debris corresponding
to each of its pericentric approaches to the Milky Way.
We adopt the LM10  scheme in which tidal debris is color-coded according to the pair of orbits on 
which it became unbound from the Sgr core: Black points are currently bound to Sgr,
orange points became unbound in the last 1.3 Gyr, magenta points between 1.3 and 3.2 Gyr ago, cyan points between 3.2 and 5.0 Gyr ago, 
and green points between 5.0 and 6.9 Gyr ago (see Fig. 7 of LM10).

Since the Sgr streams frequently overlap themselves in angular position, distance, and radial velocity, it is important to distinguish the individual wraps of these streams from each other
when evaluating possible associations with stellar subsystems in the Galactic halo.
We therefore follow LM10 in adopting a shorthand notation describing specific wraps of the streams (see also Fig. 8 of LM10):

\begin{description}

\item[L1:]  Primary wrap of the leading arm of Sgr (i.e., $0^{\circ} - 360^{\circ}$ angular separation from the dwarf).  Traced by orange/magenta points near the Sgr dwarf, magenta/cyan/green points at larger angular separations.

\item[L1Y:] Young subset of the L1 wrap, incorporating only orange/magenta points (i.e., debris lost during the last 3 Gyr).

\item[L2:] Secondary wrap of the leading arm of Sgr (i.e., $360^{\circ} - 720^{\circ}$ angular separation from the dwarf).  Traced almost entirely by cyan/green points.

\item[T1:]  Primary wrap of the trailing arm of Sgr (i.e., $0^{\circ} - 360^{\circ}$ angular separation from the dwarf).  Traced by orange/magenta points near the Sgr dwarf, magenta/cyan/green points at larger angular separations.

\item[T1Y:] Young subset of the T1 wrap, incorporating only orange/magenta points (i.e., debris lost during the last 3 Gyr).

\item[T2:] Secondary wrap of the trailing arm of Sgr (i.e., $360^{\circ} - 720^{\circ}$ angular separation from the dwarf).  Traced almost entirely by cyan/green points.

\end{description}

The L1Y and T1Y stream segments described above correspond most closely to 
those phases of tidal debris $\lesssim 3$ Gyr in age for which there is the strongest observational evidence from the 2MASS and SDSS wide-field surveys.
The true length of the debris streams is unknown however, and it is possible that they may extend substantially further than has been conclusively observed to date.
The L2/T2 streams (and the cyan/green sections of L1/T1) represent model predictions for where such putative older
tidal debris may lie, should it be present in the Galactic halo.

\section{SELECTING A SATELLITE POPULATION}
\label{gcsel.sec}

As illustrated in Figure \ref{XYfig.fig}, the most recent (i.e., $\lesssim 3$ Gyr in age) Sgr tidal debris
lies at distances $10 \lesssim d \lesssim 60$ kpc from the Sun, and greater than 8 kpc from the Galactic Center.
The LM10 simulations suggest, however, that putative L2/T2 tidal debris may exist as much as 100 kpc away.
We therefore construct our comparison sample of Galactic globular clusters by extracting all those
with Galactocentric radii $8 < r_{\rm GC} < 100$ kpc from the Harris (1996, with Feb. 2003 update)
catalog; this conservatively includes all well-known globular clusters that might conceivably be associated with Sgr.
We note that this rules out checking for possible angular alignments with much more distant clusters that may have been lost extremely at extremely early times;
we discuss this point further in \S \ref{nearfargcs.sec}.
Motivated by claims for association with Sgr tidal debris by Carraro et al. (2007, 2009), we also include in our sample
the star clusters Whiting 1, Berkeley 29 (Be 29), and Saurer 1 (Sa 1).  While Be 29 and Sa 1 are commonly regarded as Galactic open clusters, their physical properties
overlap with the lower-mass globular clusters in our sample, and we consider that they may be sparse halo globular clusters which happen to lie near the Galactic disk.

Recent years have witnessed the discovery in the Sloan Digital Sky Survey (SDSS) of a multitude  of ``ultra-faint'' 
Galactic satellites  that have significantly lower luminosity than the classic dwarf spheroidal galaxy population, but half-light radii too large
to be traditional globular clusters (see, e.g., discussion by Gilmore et al. 2007).
The origin of this low luminosity population is uncertain, and some have been suggested to be stellar overdensities
previously associated at some time in the past with massive dSphs such as Sgr.
We therefore include in our analysis the ultra-faint systems Bo\"{o}tes I (Boo I), Bo\"{o}tes II (Boo II), 
Coma Berenices (Coma Ber), Pisces I, Segue 1, Segue 2, Segue 3, Ursa Major II (UMa II), and Willman I.
The Canes Venatici I, Canes Venatici II, Hercules, Leo IV, Leo V, Leo T, and Ursa Major I systems along with the classical dSph population
(e.g., Carina, Draco, Fornax, etc.) are excluded from formal consideration
since they lie at large distances (typically greater than 100 kpc)  at which there is no Sgr tidal debris in the LM10 model, although we discuss some
of them briefly in \S \ref{distantUFs.sec}.

\begin{deluxetable*}{rlrcccccccccc}
\tablecolumns{13}
\tablewidth{0pc}
\tabletypesize{\scriptsize}
\tablecaption{Properties of the Milky Way Satellites}
\tablehead{\colhead{ID} & \colhead{Name} & \colhead{R.A.} & \colhead{Decl.} & \colhead{$l$} & \colhead{$b$} & \colhead{$| Z_{\rm Sgr} |$} & \colhead{$\Lambda_{\odot}$} & 
\colhead{$B_{\odot}$\tablenotemark{a}} & \colhead{Distance} & \colhead{$V_{\rm GSR}$} & \colhead{$\mu_{\alpha} {\textrm cos}\delta$} & 
\colhead{$\mu_{\delta}$} \\ 
\colhead{} & \colhead{} & \colhead{(J2000.0)} & \colhead{(J2000.0)} & \colhead{($^{\circ}$)} & \colhead{($^{\circ}$)} & \colhead{(kpc)} & \colhead{($^{\circ}$)} & \colhead{($^{\circ}$)} &
\colhead{(kpc)} & \colhead{(km s$^{-1}$)} & \colhead{(mas yr$^{-1}$)} &  \colhead{(mas yr$^{-1}$)}}
\startdata
\multicolumn{13}{c}{Star Clusters}\\
\hline
1	 & 	NGC 288	 & 		00:52:48	 & 	-26:35:24	 & 152 & -89 & 	1.1	 & 	76	 & 	13	 & 	8.8	 & 	$-52.5\pm0.4$	 & 	$4.68\pm0.18$	 & 	$-5.37\pm0.52$ \\
2	 & 	NGC 362	 & 		01:03:14	 & 	-70:50:54	 & 302 & -46 & 	5.59	 & 	46	 & 	50	 & 	8.5	 & 	$85\pm0.5$	 & 	$5.08\pm1.15$	 & 	$-2.49\pm2.17$ \\
3	 & 	Whiting 1	 & 		02:02:57	 & 	-03:15:10	 & 162 & -61 &	0.215	 & 	103	 & 	1	 & 	29.4	 & 	$-105\pm1.8$	 & 	...	 & 	... \\
4	 & 	NGC 1261	 & 		03:12:15	 & 	-55:13:01	 & 271 & -52 &	11.9	 & 	79	 & 	51	 & 	16.4	 & 	$-79.7\pm4.6$	 & 	...	 & 	... \\
5	 & 	Pal 1	 & 			03:33:23	 & 	+79:34:50	 & 130 & 19 &	9.8	 & 	180	 & 	-55 & 	10.9	 & 	$81.9\pm3.3$	 & 	...	 & 	... \\
6	 & 	Eridanus	 & 		04:24:45	 & 	-21:11:13	 & 218 & -41 &	50.1	 & 	125	 & 	35	 & 	90.2	 & 	$-141\pm2.1$	 & 	...	 & 	... \\
7	 & 	Pal 2	 & 			04:46:06	 & 	+31:22:51	 & 171 & -9 &	5.95	 & 	154	 & 	-11	 & 	27.6	 & 	$-105\pm57$	 & 	...	 & 	... \\
8	 & 	NGC 1851	 & 		05:14:06	 & 	-40:02:50	 & 245 & -35 &	9.13	 & 	121	 & 	56	 & 	12.1	 & 	$142\pm0.6$	 & 	$1.28\pm0.68$	 & 	$2.39\pm0.65$ \\
9	 & 	M 79 (NGC 1904) & 	05:24:11 & 	-24:31:27	 & 227 & -29 &	8.1	 & 	138	 & 	44	 & 	12.9	 & 	$48.8\pm0.4$	 & 	$2.12\pm0.64$	 & 	$-0.02\pm0.64$ \\
10	 & 	NGC 2298	 & 		06:48:59	 & 	-36:00:19	 & 246 & -16 &	8.61	 & 	157	 & 	63	 & 	10.7	 & 	$-59.7\pm1.2$	 & 	$4.05\pm1$	 & 	$-1.72\pm0.98$ \\
11	 & 	Berkeley 29 & 		06:53:04	 & 	+16:55:39	 & 198 & 8 &	1.79	 & 	177	 & 	12	 & 	13.2	 & 	$-53.7\pm3.6$	 & 	...	 & 	... \\
12	 & 	Saurer 1	 & 		07:20:54	 & 	+01:48:00	 & 215 & 7 &	5.25	 & 	182	 & 	28	 & 	13.2	 & 	$-32.8\pm3.6$	 & 	...	 & 	... \\
13	 & 	NGC 2419	 & 		07:38:09	 & 	+38:52:55	 & 180 & 25 &	13.6	 & 	190	 & 	-9	 & 	84.2	 & 	$-26.5\pm0.8$	 & 	...	 & 	... \\
14	 & 	Pyxis	 & 		09:07:58	 & 	-37:13:17	 & 261 & 7 &	35.4	 & 	224	 & 	66	 & 	39.7	 & 	$-194\pm1.9$	 & 	...	 & 	... \\
15	 & 	NGC 2808	 & 		09:12:03	 & 	-64:51:47	 & 282 & -11 &	8.6	 & 	332	 & 	82	 & 	9.6	 & 	$-128\pm2.4$	 & 	$0.58\pm0.45$	 & 	$2.06\pm0.46$ \\
16	 & 	Pal 3	 & 			10:05:31	 & 	+00:04:17	 & 240 & 42 &	40.6	 & 	228	 & 	27	 & 	92.7	 & 	$-65.1\pm8.4$	 & 	$0.33\pm0.23$	 & 	$0.3\pm0.31$ \\
17	 & 	NGC 3201	 & 		10:17:37 & 	-46:24:40	 & 277 & 9 &	3.73	 & 	265	 & 	68	 & 	5	 & 	$269\pm0.2$	 & 	$5.28\pm0.32$	 & 	$-0.98\pm0.33$ \\
18	 & 	NGC 4147	 & 		12:10:06	 & 	+18:32:31	 & 253 & 77 &	1.39	 & 	251	 & 	-1 & 	19.3	 & 	$140\pm0.7$	 & 	$-1.63\pm1.44$	 & 	$-1.86\pm3.21$ \\
19	 & 	Rup 106	 & 		12:38:40 & 	-51:09:01	 & 301 & 12 &	16.1	 & 	304	 & 	53	 & 	21.2	 & 	$-233\pm3$	 & 	...	 & 	... \\
20	 & 	M 68 (NGC 4590) & 	12:39:28	 & 	-26:44:34	 & 300 & 36 &	4.92	 & 	281	 & 	35	 & 	10.2	 & 	$-250\pm0.4$	 & 	$-3.76\pm0.66$	 & 	$1.79\pm0.62$ \\
21	 & 	M 53 (NGC 5024) & 	13:12:55 & 	+18:10:09	 & 333 & 80 &	3.4	 & 	265	 & 	-8	 & 	17.8	 & 	$-89.5\pm4.1$	 & 	$0.5\pm1$	 & 	$-0.1\pm1$ \\
22	 & 	NGC 5053	 & 		13:16:27	 & 	+17:41:53	 & 336 & 79 &	3.21	 & 	266	 & 	-8	 & 	16.4	 & 	$34.1\pm0.4$	 & 	...	 & 	... \\
23	 & 	M 3 (NGC 5272)	 & 	13:42:11 & 	+28:22:32	 & 42 & 79 &	4.5	 & 	265	 & 	-20	 & 	10.4	 & 	$-109\pm0.2$	 & 	$-2.84\pm2.01$	 & 	$-2.45\pm0.62$ \\
24	 & 	NGC 5286	 & 		13:46:27 & 	-51:22:24	 & 312 & 11 &	6.94	 & 	315	 & 	45	 & 	11	 & 	$-106\pm1.5$	 & 	...	 & 	... \\
25	 & 	AM 4	 & 			13:55:50 & 	-27:10:22	 & 320 & 34 &	12	 & 	298	 & 	26	 & 	29.9	 & 	...	 & 	...	 & 	... \\
26	 & 	NGC 5466	 & 		14:05:27	 & 	+28:32:04	 & 42 & 74 &	7.16	 & 	270	 & 	-23	 & 	15.9	 & 	$160\pm0.3$	 & 	$-4.46\pm1.32$	 & 	$0.85\pm0.84$ \\
27	 & 	NGC 5634	 & 		14:29:37 & 	-05:58:35	 & 342 & 49 &	0.531	 & 	293	 & 	3	 & 	25.2	 & 	$-80.5\pm6.6$	 & 	...	 & 	... \\
28	 & 	NGC 5694	 & 		14:39:37 & 	-26:32:18	 & 331 & 30 &	10.7	 & 	306	 & 	20	 & 	34.7	 & 	$-231\pm1.3$	 & 	...	 & 	... \\
29	 & 	IC 4499	 & 		15:00:19	 & 	-82:12:49	 & 307 & -20 &	15	 & 	2.89	 & 	57	 & 	18.9	 & 	...	 & 	...	 & 	... \\
30	 & 	NGC 5824	 & 		15:03:59	 & 	-33:04:04	 & 333 & 22 &	11.2	 & 	315	 & 	22	 & 	32	 & 	$-117\pm1.5$	 & 	...	 & 	... \\
31	 & 	Pal 5	 & 			15:16:05	 & 	+00:06:41	 & 1 & 46 &	4.06	 & 	300	 & 	-8 & 	23.2	 & 	$-44.3\pm0.2$	 & 	$-2.29\pm0.67$	 & 	$-1.65\pm0.81$ \\
32	 & 	BH 176	 & 		15:39:07	 & 	-50:03:02	 & 328 & 4 &	7.58	 & 	332	 & 	33	 & 	15.6	 & 	...	 & 	...	 & 	... \\
33	 & 	Pal 14	 & 		16:11:05	 & 	+14:57:29	 & 29 & 42 &	35	 & 	305	 & 	-27	 & 	73.9	 & 	$170\pm2.2$	 & 	...	 & 	... \\
34	 & 	NGC 6101	 & 		16:25:49 & 	-72:12:06	 & 318 & -16 &	10.4	 & 	355	 & 	47	 & 	15.3	 & 	$216\pm1.7$	 & 	...	 & 	... \\
35	 & 	M 13 (NGC 6205)  & 	16:41:42 & 	+36:27:37	 & 59 & 41 &	6.72	 & 	297	 & 	-49	 & 	7.7	 & 	$-87.2\pm0.3$	 & 	$-0.9\pm0.71$	 & 	$5.5\pm0.89$ \\
36	 & 	NGC 6229	 & 		16:46:59 & 	+47:31:40	 & 74 & 40 &	26.6	 & 	285	 & 	-58	 & 	30.4	 & 	$22\pm7.6$	 & 	...	 & 	... \\
37	 & 	Pal 15	 & 		17:00:02	 & 	+00:32:31	 & 20 & 25 &	16.2	 & 	324	 & 	-20 & 	44.6	 & 	$151\pm1.1$	 & 	...	 & 	... \\
38	 & 	M 92 (NGC 6341) & 	17:17:07	 & 	+43:08:11	 & 68 & 35 &	7.9	 & 	298	 & 	-59	 & 	8.2	 & 	$63.4\pm0.1$	 & 	$-3.75\pm1.72$	 & 	$0.39\pm1.68$ \\
39	 & 	IC 1257	 & 		17:27:09	 & 	-07:05:35	 & 17 & 15 &	7.61	 & 	334	 & 	-16	 & 	25	 & 	$-66.3\pm2.1$	 & 	...	 & 	... \\
40	 & 	NGC 6426	 & 		17:44:55 & 	+03:10:13	 & 28 & 16 &	10.2	 & 	334	 & 	-27	 & 	20.7	 & 	$-47.5\pm23$	 & 	...	 & 	... \\
41	 & 	ESO 280-SC06	 & 	18:09:06	 & 	-46:25:23	 & 347 & -13 &	6.17	 & 	355	 & 	19	 & 	21.7	 & 	...	 & 	...	 & 	... \\
42	 & 	M 54 (NGC 6715) & 	18:55:03	 & 	-30:28:42	 & 6 & -14 &	0.199	 & 	0	 & 	1	 & 	28.0	 & 	$171\pm0.5$	 & 	...	 & 	... \\
43	 & 	M 56 (NGC 6779) & 	19:16:36 & 	+30:11:05	 & 63 & 8 &	9.56	 & 	352	 & 	-59 & 	10.1	 & 	$73.3\pm0.8$	 & 	$0.3\pm1$	 & 	$1.4\pm1$ \\
44	 & 	Terzan 7	 & 		19:17:44 & 	-34:39:27	 & 3 & -20 &	1.09	 & 	5	 & 	5	 & 	23.2	 & 	$185\pm4$	 & 	...	 & 	... \\
45	 & 	Arp 2	 & 			19:28:44 & 	-30:21:14	 & 9 & -21 &	0.71	 & 	7	 & 	0	 & 	28.6	 & 	$153\pm10$	 & 	...	 & 	... \\
46	 & 	Terzan 8	 & 		19:41:45	 & 	-34:00:01	 & 6 & -25 &	0.812	 & 	10	 & 	4	 & 	26	 & 	$156\pm8$	 & 	...	 & 	... \\
47	 & 	M 75 (NGC 6864) & 	20:06:05	 & 	-21:55:17	 & 20 & -26 &	3.97	 & 	15	 & 	-9 & 	20.7	 & 	$-112\pm3.6$	 & 	...	 & 	... \\
48	 & 	NGC 6934	 & 		20:34:12 & 	+07:24:15	 & 52 & -19 &	10.5	 & 	23	 & 	-38	 & 	15.7	 & 	$-235\pm1.6$	 & 	$1.2\pm1$	 & 	$-5.1\pm1$ \\
49	 & 	M 72 (NGC 6981) & 	20:53:28 & 	-12:32:13	 & 35 & -33 &	6.01	 & 	27	 & 	-18 & 	17	 & 	$-230\pm3.7$	 & 	...	 & 	... \\
50	 & 	NGC 7006	 & 		21:01:30	 & 	+16:11:15	 & 64 & -19 &	30.7	 & 	33	 & 	-46	 & 	41.5	 & 	$-186\pm0.4$	 & 	$-0.96\pm0.35$	 & 	$-1.14\pm0.4$ \\
51	 & 	M 15 (NGC 7078) & 	21:29:58 & 	+12:10:01	 & 65 & -27 &	7.6	 & 	42	 & 	-41	 & 	10.3	 & 	$80\pm0.2$	 & 	$-1.28\pm1.34$	 & 	$-7.05\pm3.09$ \\
52	 & 	M 2 (NGC 7089)	 & 	21:33:29 & 	+00:49:23	 & 55 & -35 &	6.53	 & 	40	 & 	-29	 & 	11.5	 & 	$151\pm2$	 & 	$6.03\pm0.99$	 & 	$-5.19\pm1.34$ \\
53	 & 	Pal 12	 & 		21:46:39 & 	-21:15:03	 & 31 & -48 &	3.24	 & 	39	 & 	-7 & 	19.1	 & 	$107\pm1.5$	 & 	$-1.2\pm0.3$	 & 	$-4.21\pm0.29$ \\
54	 & 	Pal 13	 & 		23:06:44	 & 	+12:46:19	 & 87 & -43 &	15.1	 & 	70	 & 	-34	 & 	25.8	 & 	$190\pm0.5$	 & 	$2.3\pm0.26$	 & 	$0.27\pm0.25$ \\
55	 & 	NGC 7492	 & 	23:08:27	 & 	-15:36:41	 & 53 & -63 &	4.06	 & 	59	 & 	-7	 & 	25.8	 & 	$-128\pm4.3$	 & 	...	 & 	... \\
\hline
\multicolumn{13}{c}{Ultrafaint Dwarfs}\\
\hline
56	 & 	Segue 2	 & 	02:19:16	 & 	+20:10:31	 & 149 & -38 &	11	 & 	119	 & 	-17	 & 	35	 & 	$43.2\pm2.5$	 & 	...	 & 	... \\
57	 & 	UMa II	 & 	08:51:30	 & 	+63:07:48	 & 152 & 37 &	18.3	 & 	202	 & 	-33	 & 	32	 & 	$-33.4\pm1.9$	 & 	...	 & 	... \\
58	 & 	Segue 1	 & 	10:07:03	 & 	+16:04:25	 & 220 & 50 &	3.5	 & 	225	 & 	11	 & 	23	 & 	$111\pm1.3$	 & 	...	 & 	... \\
59	 & 	Willman I	 & 	10:49:22	 & 	+51:03:04	 & 159 & 57 &	16.9	 & 	223	 & 	-25	 & 	38	 & 	$35.4\pm2.5$	 & 	...	 & 	... \\
60	 & 	Coma Ber	 & 	12:26:59	 & 	+23:54:15	 & 242 & 84 &	7.03	 & 	252	 & 	-8	 & 	44	 & 	$81.8\pm0.9$	 & 	...	 & 	... \\
61	 & 	Bootes II	 & 	13:58:05	 & 	+12:52:00	 & 354 & 69 &	8.09	 & 	277	 & 	-9	 & 	46	 & 	$-116\pm5.2$	 & 	...	 & 	... \\
62	 & 	Bootes I	 & 	14:00:06	 & 	+14:30:00	 & 358 & 70 &	12.4	 & 	277	 & 	-11	 & 	62	 & 	$103\pm3.4$	 & 	...	 & 	... \\
63	 & 	Segue 3	 & 	21:21:31	 & 	+19:07:02	 & 69 & -21 &	12.8	 & 	41	 & 	-48	 & 	16	 & 	...	 & 	...	 & 	... \\
64	 & 	Pisces I	 & 	23:40:00	 & 	+00:18:00	 & 88 & -58 &	28.2	 & 	73	 & 	-19	 & 	85	 & 	$45.4\pm4$	 & 	...	 & 	... \\
\enddata
\label{gcprops.table}
\end{deluxetable*}

The final sample of 64 satellites is tabulated in Table \ref{gcprops.table} (sorted in order of increasing right ascension);
59 of these satellites have both distance and radial velocity measurements published in the literature.  Of these 59 satellites, 51 are stellar clusters
and eight are ultrafaint dwarfs.
Proper motions\footnote{Because the apparent proper motion of the Sgr stream along the direction of Galactic longitude $l$  varies extremely rapidly
for tidal debris passing near the poles of the Galactic $(l,b)$ coordinate system we opt to present proper motions in the $(\alpha,\delta)$ coordinate frame.}
have been measured for 24 of these satellites, and  have been drawn from the Harris (1996) catalog,  
Dinescu et al. (1999, 2001, 2003), Siegel et al. (2001), Palma et al. (2002), and Cassetti-Dinescu et al. (2007).


\section{QUANTIFYING THE ASSOCIATION OF SATELLITES }
\label{quantifying.sec}

While the best evaluation of the association of any individual satellite can be made by considering the properties of the satellite in comparison
to those of the Sgr stream in detail, a preliminary cut of reasonable stream candidates can be made via simple statistical
arguments.  In this section, we describe our construction of a correlation statistic that can be used to pre-select those Milky Way satellites that
match the Sgr streams sufficiently well to warrant close inspection in \S \ref{compareGC.sec} and \ref{compareUF.sec}.


\subsection{Defining an Association Statistic}
We quantify the offset of each satellite listed in Table \ref{gcprops.table} 
from the LM10 model of the Sgr stream as the quadrature sum of the angular separation, 
difference in heliocentric distance, and difference in radial velocity between the satellite and the stream:
\begin{equation}
\chi^2 = \frac{(B_{\odot} - B_{\odot,{\rm Sgr}})^2}{\sigma_{B_{\odot,{\rm Sgr}}}^2} +  \frac{(d - d_{\rm Sgr})^2}{\sigma_{\rm d, Sgr}^2 + \sigma_{\rm d}^2} + \frac{(v - v_{\rm Sgr})^2}{\sigma_{\rm v, Sgr}^2 + \sigma_{\rm v}^2}
\end{equation}
where $B_{\odot}$, $d$, and $v$ are the latitude (in Sgr plane coordinates), the heliocentric distance, and the radial 
velocity of the satellite respectively.  
$B_{\odot,{\rm Sgr}}$, $\sigma_{B_{\odot,{\rm Sgr}}}$, $d_{\rm Sgr}$, $\sigma_{\rm d, Sgr}$, $v_{\rm Sgr}$, 
and $\sigma_{\rm v, Sgr}$ respectively represent the mean and $1\sigma$ width of the latitude, distance, and radial velocity
of the LM10 stream, calculated within $5^{\circ}$ of orbital longitude $\Lambda_{\odot}$ of the satellite
(i.e., taking $\Lambda_{\odot}$ as the independent variable).
We assume a typical observational uncertainty $\sigma_{\rm d} / d = 10\%$, and $\sigma_{\rm v} = 3$ \kms for each of the satellites.
$\chi^2$ therefore quantifies the offset between a satellite and the Sgr stream, weighted by the quadrature sum of the observational uncertainty and the
variance of Sgr stream stars about the mean
at the corresponding location.
$\chi^2$ is calculated with respect to each of the arms of the Sgr stream listed above in \S \ref{model.sec} (i.e., $\chi_{\rm L1}^2$, $\chi_{\rm L1Y}^2$, $\chi_{\rm L2}^2$, etc.).

\subsubsection{Application to the Globular Clusters}
\label{gcstats.sec}

Taken on its own, $\chi^2$ is  a marginally useful statistic that quantifies how closely
a given satellite matches the properties of the Sgr stream.  A more interesting question, however, is the following:
How significant is the apparent association compared to that expected for a population of satellites randomly distributed throughout the Galactic halo?
To assess this, we 
perform a series of
Monte Carlo simulations, randomly populating the Galactic halo with $10^4$ artificial satellites between Galactocentric radii
8 kpc  $< r_{\rm GC} <$ 100 kpc.  We assume a spherically symmetric distribution with a radial density profile proportional
to $r_{\rm GC}^{-1.6}$ (Bellazzini et al. 2003a), which mimics the Galactic halo globular cluster population in this range of Galactocentric radii.
These physical locations are transformed to the heliocentric $(\Lambda_{\odot}, B_{\odot}, d)$ coordinate frame,
and paired with a radial velocity drawn at random from a Gaussian distribution with mean 
$\langle v_{\rm GSR} \rangle = -38$ \kms and standard deviation $\sigma_{v_{\rm GSR}} = 175$ \kms
(see Bellazzini et al. 2003a).\footnote{Adopting $\langle v_{\rm GSR} \rangle = 0$ \kms instead makes a negligible difference to our calculations.}  

Calculating $\chi^2$ for each of these $10^4$ artificial globular clusters, we determine the fraction $P(\chi)$ that
by chance match some arm of the Sgr stream to an accuracy of $\chi$ or better.
Since the likelihood of matching some segment of the Sgr stream depends strongly on the length of the stream (if the streams were arbitrarily long 
they would fill the available parameter space, and {\it any} randomly distributed satellite would match {\it some} wrap reasonably well)
we calculate two versions of the $P$ statistic:

\begin{description}

\item[$P_3$:] Fraction of randomly distributed  satellites that match either the L1Y or T1Y arms (i.e., the last 3 Gyr of tidal debris) to an accuracy better than $\chi$.

\item[$P_7$:] Fraction of randomly distributed  satellites that match either the L1, L2, T1, or T2 arms (i.e., the last 7 Gyr of tidal debris) to an accuracy better than $\chi$.

 \end{description}

\begin{figure}
\plotone{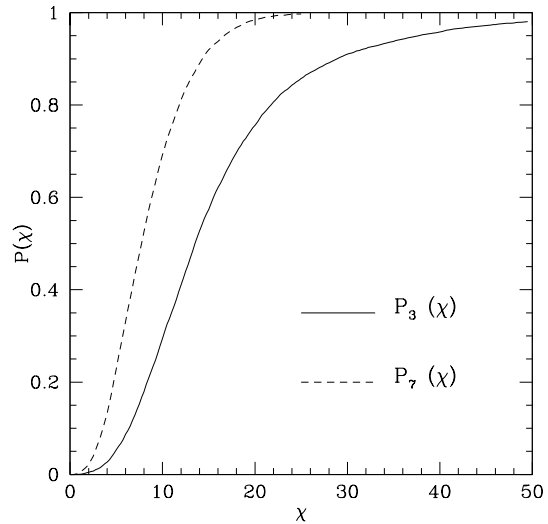}
\caption{Probability $P(\chi)$ for an artificial halo globular cluster to
match the Sgr stream with an accuracy of $\chi$ or better.
The solid line is the relation obtained when considering only matches to the most recent 3 Gyr of tidal debris (i.e., L1Y and T1Y wraps), the dashed line is the relation obtained when
considering matches to the most recent 7 Gyr of tidal debris (i.e., L1, T1, L2, and T2 wraps).}
\label{Random.fig}
\end{figure}

If Sgr has only been interacting with the Milky Way for the last 3 Gyr (i.e., there is no significant tidal debris beyond that conclusively traced to date by 2MASS and SDSS)
the $P_3$ statistic will be most realistic, whereas if there is appreciable tidal debris from earlier epochs not traced by current surveys then
the $P_7$ statistic will be most appropriate.  We plot $P_3$ and $P_7$ as functions of $\chi$ in Figure \ref{Random.fig}, noting that $P_3$ and $P_7$ can also be interpreted as
the {\it probability} that some randomly chosen artificial globular cluster will match the Sgr streams to an accuracy of $\chi$ or better.
Figure \ref{Random.fig} illustrates that there is a low probability for a randomly chosen artificial cluster to 
match the Sgr stream to high accuracy ($\chi \lesssim 3$), while nearly all artificial clusters will match the stream to better than
$\chi \sim 30$.  Unsurprisingly, $P_7$ rises faster than $P_3$ as a function of $\chi$ because there are more opportunities for each cluster to match some wrap of the tidal streams.

Using the relations shown in Figure \ref{Random.fig}, it is possible to calculate the probabilities
$P_{\rm 7, L1}$, $P_{\rm 3, L1Y}$, etc., that a randomly chosen artificial
cluster would match some segment of the Sgr stream as well as (or better than) the real cluster matches a given segment of the stream.
Values of these $P$ are tabulated for each of our 51 globular clusters with full angular, distance, and radial velocity information in Tables
\ref{results1.table} and \ref{results2.table}.  
Note that  $P_3$ is defined as the minimum of  $P_{\rm 3, L1Y}$ and $P_{\rm 3, T1Y}$,
and $P_7$ as the minimum of $P_{\rm 7, L1}$, $P_{\rm 7, L2}$, $P_{\rm 7, T1}$, and $P_{\rm 7, T2}$.
$P_3$ and $P_7$ therefore summarize the significance of the association of a given cluster with the Sgr streams under two different assumptions for the length
of the streams.

\begin{figure}
\plotone{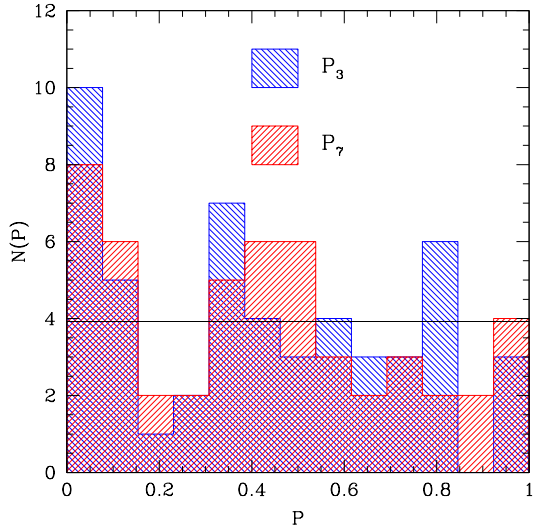}
\caption{Number of globular clusters as a function of the probability statistic $P$.  A randomly distributed cluster population would
be uniform from $P = 0 -1$ (solid horizontal line) modulo statistical fluctuations, while the actual Milky Way distribution exhibits a peak at $P \leq 0.15$ 
on account of the presence of 
clusters genuinely associated with the Sgr stream.  The red histogram corresponds to association probabilities considering all wraps
of the LM10 model stream ($P_7$), the blue histogram indicates the result when only dynamically young sections most closely corresponding to observational
data are used ($P_3$).}
\label{nofp.fig}
\end{figure}

In Figure \ref{nofp.fig} we plot a histogram of $P_3$ and $P_7$ for our 51 globular clusters.  If these globular clusters were distributed randomly throughout the Galactic halo, 
the number $N(P)$ of satellites in each bin would be uniform throughout the range $P = 0 - 1$ (modulo statistical fluctuations).  
The spike in the number of clusters with $P \leq 0.15$ reflects the fact that there is a globular cluster population genuinely associated with the Sgr tidal
streams; we therefore adopt the criterion $P_3 \leq 0.15$ or $P_7 \leq 0.15$ for selecting globular clusters as candidates for association with Sgr.
Based on our Monte Carlo simulations of $10^4$ randomly distributed artificial clusters, and 500 realizations of randomly extracted sets of 51 such clusters, 
we expect that $7 \pm 2$ globular clusters will meet the selection criteria (either $P_3$ or $P_7$) based 
on chance alignment with the Sgr streams.
In comparison, 15/14 globular clusters actually meet the $P_3$/$P_7$
criteria respectively.\footnote{Note that some clusters that are Sgr stream candidates using the $P_3$ statistic 
are not candidates under the $P_7$ statistic because by opening the parameter space for
potential matches to include alignment with the older arms, the Monte Carlo probability normalization necessarily imposes more stringent constraints on possible associations with
the young arms.}
In less than 1\% of realizations of a sample of 51 randomly drawn artificial clusters do we find such a large number with $P_3$ or $P_7 \leq 0.15$ by chance.
Subtracting from the maximum number of potential Sgr globular clusters (15) the typical number expected to be false associations ($7 \pm 2$), we conclude that there
are likely $8 \pm 2$ globular clusters genuinely associated with the Sgr stream.


%

\subsubsection{Application to the Ultrafaint Satellites}

The $P$ statistic defined in \S \ref{gcstats.sec} above is not appropriate for the ultra-faint satellite population discovered in the SDSS.
Not only do these satellites have a different radial number density profile than the globular clusters, but they are subject to a strong selection bias
in the sense that only satellites lying within the SDSS footprint could have been discovered.
Since the Sgr stream occupies a significant fraction of this
footprint, the ultra-faint satellites discovered to date will therefore be predisposed to lie relatively close to the stream.

\begin{figure}
\plotone{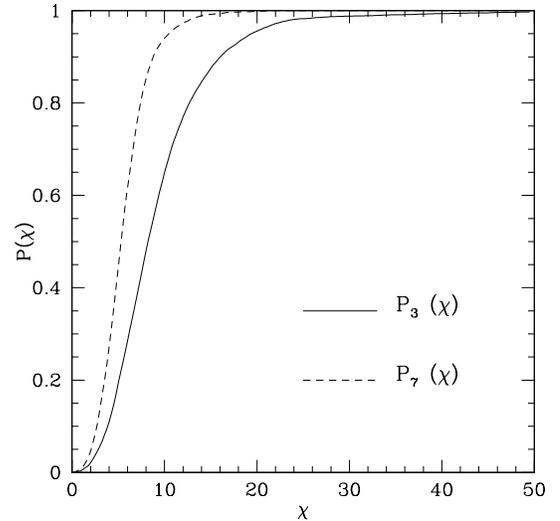}
\caption{As Figure \ref{Random.fig}, but for an artificial population constructed to mimic the distribution of Milky Way  ultra-faint satellites.}
\label{RandomUF.fig}
\end{figure}

\begin{figure}
\plotone{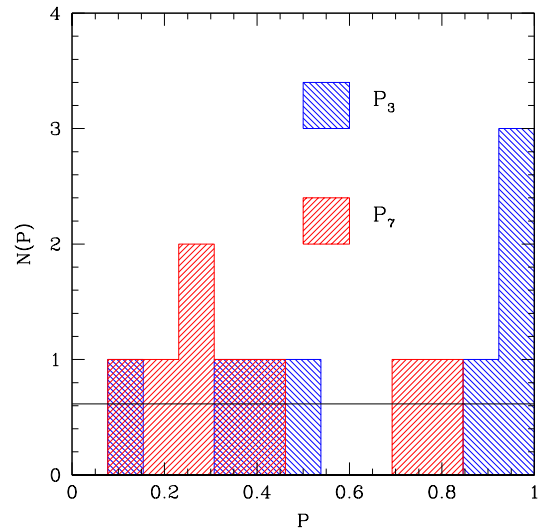}
\caption{As Figure \ref{nofp.fig}, but for the ultra-faint Milky Way satellites.}
\label{nofpuf.fig}
\end{figure}

We therefore develop an alternative formulation of the $P$ statistic appropriate to the ultra-faint satellites.  
Since the radial number density profile of this population is poorly constrained by the small number of such satellites discovered to date, 
we generate an artificial comparison population whose Galactocentric radii  match those of the observed satellites.
We assume that this artificial population is spherically symmetric about the Milky Way, but reject from consideration any satellites that fall outside
of the SDSS footprint.\footnote{We use an SDSS DR7 lookup table kindly provided by M. Juric 2010 (priv. comm.).}
Sufficiently many artificial satellites are generated that the final population that falls within the SDSS footprint is composed of $10^4$ members, for each
of which a radial velocity is chosen from a 
Gaussian random distribution with mean 
$\langle v_{\rm GSR} \rangle = 0$ \kms and standard deviation $\sigma_{v_{\rm GSR}} = 75$ \kms
that fits the observed distribution of $v_{\rm GSR}$ for the real satellites in question.

In Figure \ref{RandomUF.fig} we plot $P$ as a function of $\chi$ for this artificially generated comparison population, noting that its cumulative distribution
rises much more steeply than did the cumulative distribution of
 the artificial globular clusters (Figure \ref{Random.fig}).  This reflects the fact that since the ultra-faint satellites by definition lie within the SDSS footprint, they are
predisposed to relatively strong apparent correlations with the Sgr stream.
Figure \ref{nofpuf.fig} plots the histogram of $P_3$ and $P_7$ for the eight ultra-faint satellites whose radial velocities have been measured.
The roughly uniform distribution between $P = 0 - 1$ indicates that the satellites are statistically  consistent with being unassociated with the Sgr stream.
In particular, there is no clump at low values of $P$ corresponding to close association with the Sgr streams.  Indeed, there may be weak evidence (at the $\sim 2\sigma$ level) 
using the $P_3$
statistic for an {\it anti-correlation} between some of these satellites and the Sgr streams as indicated by the high $N(P)$ for $P_3 > 0.9$ (blue histogram in Figure \ref{nofpuf.fig}).
Caution is therefore advisable in the evaluation of individual ultra-faint systems given their propensity  to appear physically
close to the Sgr streams simply by virtue of the location of the SDSS survey fields.

\subsection{Caveats}

Our numerical method of quantifying the degree of association of individual Milky Way satellites with the Sgr stellar debris streams
is based upon some significant assumptions.
First, we assume that the LM10
model is an accurate representation of the real Sgr stream.  While this is a good assumption
for the L1Y and T1Y wraps of the leading and trailing tidal arms (where the model was constrained
to match observational data), it is unknown how well the model may describe 
tidal debris more widely separated from the dwarf.  It may be the case that the most extended regions of the tidal
tails (if they exist in the Galactic halo) depart significantly from the LM10 model.
Indeed, since the LM10 model does not explicitly include orbital evolution of the Sgr dwarf (e.g., via dynamical friction),
some mismatch in the oldest sections of the tidal streams is likely.
We therefore generally place our greatest confidence in the $P_3$ statistic, which quantifies the 
association of individual systems with only those segments of the Sgr stream that have been conclusively detected
in the Galactic halo and used to constrain the LM10 model.


Second, we assume that globular clusters stripped from Sgr 
follow the path of Sgr tidal debris.  We believe this to be a reasonable assumption because neither unbound tidal debris nor globular clusters
are sufficiently massive to have their orbits noticeably altered by dynamical friction in the gravitational potential of the Milky Way.

Third, and particularly relevant to the case of the ultra-faint satellites, we have assumed that the artificial comparison satellite distribution is spherically
symmetric about the Milky Way.  This assumption, and the ensuing $P$ statistics derived from it, may be incorrect if the real 
satellite population is distributed in a strongly non-isotropic manner.  
Such a strongly flattened distribution is not observed for the Galactic globular cluster population, but may
prove to be the case for the ultra-faint satellite population if they tend to be aligned with  the observed plane of Milky Way satellite galaxies (see, e.g.,
Zentner et al. 2005; Metz et al. 2009), or if they fell into the Galactic halo in a group with the Sgr 
dwarf (see, e.g., Li \& Helmi 2008; D'Onghia \& Lake 2009).


Finally, we assume that no Sgr clusters have been tidally destroyed over the past few Gyr.  
While this would not affect our identification of which current satellites originated in the Sgr system, it may cause us to underestimate the
total globular cluster content of the original system.
For clusters still contained within the Sgr dwarf, this is likely a reasonable assumption (see, e.g.,  Pe{\~n}arrubia et al. 2009).
Once unbound from the Sgr gravitational potential however, the validity of this assumption is unknown
since some globular clusters are {\it known} to be tidally disrupting  in the Galactic potential (e.g., Pal 5; Odenkirchen et al. 2001),
even at significant distances from the Galactic center.

\section{ANALYSIS OF INDIVIDUAL STAR CLUSTERS}
\label{compareGC.sec}

\begin{deluxetable*}{rlcccccccc}
\tablecolumns{10}
\tablewidth{0pc}
\tabletypesize{\scriptsize}
\tablecaption{Possible Matches to the Sgr Stream\tablenotemark{a}}
\tablehead{\colhead{ID} & \colhead{Name} & \colhead{$P_3$} & \colhead{$P_7$} & \colhead{$P_{\rm 3, T1Y}$}  & \colhead{$P_{\rm 3, L1Y}$}  & \colhead{$P_{\rm 7, T1}$}  &
\colhead{$P_{\rm 7, T2}$}   & \colhead{$P_{\rm 7, L1}$} & \colhead{$P_{\rm 7, L2}$}}
\startdata
\multicolumn{10}{c}{Star Clusters}\\
\hline
1	 & 	NGC 288	 & 	{\bf 0.095}	 & 	0.315	 & 	{\bf 0.095}	 & 	0.934	 & 	0.315	 & 	0.727	 & 	0.963	 & 	0.366 \\
3	 & 	Whiting 1	 & 	{\bf 0.001}	 & 	{\bf 0.010}	 & 	{\bf 0.001}	 & 	0.796	 & 	{\bf 0.010}	 & 	0.995	 & 	0.588	 & 	{\bf 0.064} \\
7	 & 	Pal 2	 & 	{\bf 0.123}	 & 	0.316	 & 	{\bf 0.123}	 & 	0.971	 & 	0.316	 & 	0.946	 & 	0.327	 & 	0.659 \\
11	 & 	Berkeley 29	 & 	{\bf 0.027}	 & 	{\bf 0.039}	 & 	0.378	 & 	{\bf 0.027}	 & 	0.750	 & 	0.743	 & 	{\bf 0.039}	 & 	0.608 \\
18	 & 	NGC 4147	 & 	{\bf 0.058}	 & 	{\bf 0.132}	 & 	{\bf 0.058}	 & 	0.317	 & 	{\bf 0.132}	 & 	0.999	 & 	0.702	 & 	0.240 \\
21	 & 	M 53	 & 	{\bf 0.106}	 & 	{\bf 0.112}	 & 	{\bf 0.106}	 & 	0.185	 & 	{\bf 0.147}	 & 	0.805	 & 	0.422	 & 	{\bf 0.112} \\
22	 & 	NGC 5053	 & 	{\bf 0.040}	 & 	{\bf 0.032}	 & 	{\bf 0.040}	 & 	0.274	 & 	{\bf 0.032}	 & 	0.999	 & 	0.591	 & 	0.442 \\
23	 & 	M 3	 & 	0.344	 & 	{\bf 0.122}	 & 	0.412	 & 	0.344	 & 	0.412	 & 	0.732	 & 	0.690	 & 	{\bf 0.122} \\
27	 & 	NGC 5634	 & 	{\bf 0.004}	 & 	{\bf 0.031}	 & 	{\bf 0.004}	 & 	0.535	 & 	{\bf 0.031}	 & 	0.589	 & 	0.712	 & 	0.608 \\
31	 & 	Pal 5	 & 	{\bf 0.045}	 & 	{\bf 0.117}	 & 	{\bf 0.045}	 & 	0.202	 & 	{\bf 0.117}	 & 	0.582	 & 	0.499	 & 	0.853 \\
37	 & 	Pal 15	 & 	{\bf 0.128}	 & 	0.390	 & 	0.890	 & 	{\bf 0.128}	 & 	0.982	 & 	0.823	 & 	0.390	 & 	0.999 \\
39	 & 	IC 1257	 & 	0.695	 & 	{\bf 0.112}	 & 	0.695	 & 	0.698	 & 	0.761	 & 	0.990	 & 	0.962	 & 	{\bf 0.112} \\
42	 & 	M 54	 & 	{\bf 0.001}	 & 	{\bf 0.001}	 & 	{\bf 0.001}	 & 	0.975	 & 	{\bf 0.001}	 & 	0.794	 & 	0.410	 & 	0.514 \\
44	 & 	Terzan 7	 & 	{\bf 0.122}	 & 	{\bf 0.090}	 & 	{\bf 0.122}	 & 	0.869	 & 	{\bf 0.090}	 & 	0.764	 & 	0.549	 & 	0.458 \\
45	 & 	Arp 2	 & 	{\bf 0.001}	 & 	{\bf 0.015}	 & 	{\bf 0.001}	 & 	0.615	 & 	{\bf 0.015}	 & 	0.660	 & 	0.344	 & 	0.277 \\
46	 & 	Terzan 8	 & 	{\bf 0.001}	 & 	{\bf 0.009}	 & 	{\bf 0.001}	 & 	0.479	 & 	{\bf 0.009}	 & 	0.654	 & 	0.441	 & 	0.640 \\
53	 & 	Pal 12	 & 	{\bf 0.012}	 & 	{\bf 0.040}	 & 	{\bf 0.012}	 & 	{\bf 0.048}	 & 	{\bf 0.066}	 & 	0.865	 & 	{\bf 0.040}	 & 	0.835 \\
\hline
\multicolumn{10}{c}{Ultrafaint Dwarfs}\\
\hline
58	 & 	Segue 1	 & 	0.318	 & 	{\bf 0.113}	 & 	0.318	 & 	0.999	 & 	{\bf 0.113}	 & 	0.970	 & 	0.999	 & 	0.483 \\
61	 & 	Bootes II	 & 	{\bf 0.124}	 & 	0.239	 & 	0.918	 & 	{\bf 0.124}	 & 	0.566	 & 	0.662	 & 	0.239	 & 	0.758 \\
\enddata
\tablenotetext{a}{Probability $P$ that a randomly selected artificial satellite would match the indicated segment of the Sgr stream as well as,
or better than, the indicated satellite.  Values $P<0.15$ are indicated in bold font.}
\label{results1.table}
\end{deluxetable*}

\begin{deluxetable*}{rlcccccccc}
\tablecolumns{10}
\tablewidth{0pc}
\tabletypesize{\scriptsize}
\tablecaption{Systems Unlikely to Match the Sgr Stream\tablenotemark{a}}
\tablehead{\colhead{ID} & \colhead{Name} & \colhead{$P_3$} & \colhead{$P_7$} & \colhead{$P_{\rm 3, T1Y}$}  & \colhead{$P_{\rm 3, L1Y}$}  & \colhead{$P_{\rm 7, T1}$}  &
\colhead{$P_{\rm 7, T2}$}   & \colhead{$P_{\rm 7, L1}$} & \colhead{$P_{\rm 7, L2}$}}
\startdata
\multicolumn{10}{c}{Star Clusters}\\
\hline
2	 & 	NGC 362	 & 	0.463	 & 	0.403	 & 	0.463	 & 	0.715	 & 	0.841	 & 	0.952	 & 	0.403	 & 	0.910 \\
4	 & 	NGC 1261	 & 	0.233	 & 	0.371	 & 	0.233	 & 	0.948	 & 	0.602	 & 	0.767	 & 	0.982	 & 	0.371 \\
5	 & 	Pal 1	 & 	0.773	 & 	0.754	 & 	0.842	 & 	0.773	 & 	0.962	 & 	0.971	 & 	0.912	 & 	0.754 \\
6	 & 	Eridanus	 & 	0.495	 & 	0.482	 & 	0.495	 & 	0.999	 & 	0.500	 & 	0.999	 & 	0.755	 & 	0.482 \\
8	 & 	NGC 1851	 & 	0.708	 & 	0.552	 & 	0.834	 & 	0.708	 & 	0.996	 & 	0.820	 & 	0.552	 & 	0.736 \\
9	 & 	M 79	 & 	0.356	 & 	0.395	 & 	0.356	 & 	0.416	 & 	0.755	 & 	0.901	 & 	0.395	 & 	0.498 \\
10	 & 	NGC 2298	 & 	0.642	 & 	0.803	 & 	0.809	 & 	0.642	 & 	0.990	 & 	0.999	 & 	0.803	 & 	0.900 \\
12	 & 	Saurer 1	 & 	0.190	 & 	0.313	 & 	0.677	 & 	0.190	 & 	0.851	 & 	0.773	 & 	0.313	 & 	0.385 \\
13	 & 	NGC 2419	 & 	0.386	 & 	0.487	 & 	0.386	 & 	0.999	 & 	0.671	 & 	0.487	 & 	0.673	 & 	0.729 \\
14	 & 	Pyxis	 & 	0.612	 & 	0.722	 & 	0.999	 & 	0.612	 & 	0.986	 & 	0.856	 & 	0.722	 & 	0.769 \\
15	 & 	NGC 2808	 & 	0.962	 & 	0.992	 & 	0.999	 & 	0.962	 & 	0.999	 & 	0.999	 & 	0.999	 & 	0.992 \\
16	 & 	Pal 3	 & 	0.825	 & 	0.540	 & 	0.953	 & 	0.825	 & 	0.842	 & 	0.646	 & 	0.540	 & 	0.639 \\
17	 & 	NGC 3201	 & 	0.800	 & 	0.937	 & 	0.882	 & 	0.800	 & 	0.937	 & 	0.999	 & 	0.986	 & 	0.947 \\
19	 & 	Rup 106	 & 	0.615	 & 	0.781	 & 	0.615	 & 	0.763	 & 	0.781	 & 	0.938	 & 	0.982	 & 	0.782 \\
20	 & 	M 68	 & 	0.608	 & 	0.266	 & 	0.608	 & 	0.739	 & 	0.744	 & 	0.856	 & 	0.970	 & 	0.266 \\
24	 & 	NGC 5286	 & 	0.757	 & 	0.878	 & 	0.757	 & 	0.856	 & 	0.878	 & 	0.964	 & 	0.993	 & 	0.954 \\
25	 & 	AM 4\tablenotemark{b}	 & 	 0.367 	 & 	 0.597 	 & 	 0.437 	 & 	 0.367 	 & 	 0.605 	 & 	 0.774 	 & 	 0.597 	 & 	 0.612  \\
26	 & 	NGC 5466	 & 	0.531	 & 	0.505	 & 	0.531	 & 	0.538	 & 	0.505	 & 	0.999	 & 	0.853	 & 	0.938 \\
28	 & 	NGC 5694	 & 	0.414	 & 	0.385	 & 	0.414	 & 	0.717	 & 	0.387	 & 	0.639	 & 	0.970	 & 	0.385 \\
29	 & 	IC 4499\tablenotemark{b}	 & 	 0.980 	 & 	 0.885 	 & 	 0.980 	 & 	 0.999 	 & 	 0.999 	 & 	 0.885 	 & 	 0.992 	 & 	 0.999  \\
30	 & 	NGC 5824	 & 	0.287	 & 	0.438	 & 	0.287	 & 	0.579	 & 	0.438	 & 	0.731	 & 	0.876	 & 	0.837 \\
32	 & 	BH 176\tablenotemark{b}	 & 	 0.828 	 & 	 0.845 	 & 	 0.933 	 & 	 0.828 	 & 	 0.989 	 & 	 0.999 	 & 	 0.979 	 & 	 0.845  \\
33	 & 	Pal 14	 & 	0.353	 & 	0.572	 & 	0.968	 & 	0.353	 & 	0.935	 & 	0.919	 & 	0.572	 & 	0.999 \\
34	 & 	NGC 6101	 & 	0.978	 & 	0.931	 & 	0.999	 & 	0.978	 & 	0.967	 & 	0.931	 & 	0.999	 & 	0.961 \\
35	 & 	M 13	 & 	0.628	 & 	0.689	 & 	0.729	 & 	0.628	 & 	0.689	 & 	0.798	 & 	0.939	 & 	0.849 \\
36	 & 	NGC 6229	 & 	0.424	 & 	0.502	 & 	0.424	 & 	0.490	 & 	0.502	 & 	0.956	 & 	0.777	 & 	0.926 \\
38	 & 	M 92	 & 	0.551	 & 	0.901	 & 	0.782	 & 	0.551	 & 	0.901	 & 	0.972	 & 	0.901	 & 	0.991 \\
40	 & 	NGC 6426	 & 	0.804	 & 	0.364	 & 	0.877	 & 	0.804	 & 	0.952	 & 	0.999	 & 	0.988	 & 	0.364 \\
41	 & 	ESO 280-SC06\tablenotemark{b}	 & 	 0.872 	 & 	 0.453 	 & 	 0.882 	 & 	 0.872 	 & 	 0.453 	 & 	 0.944 	 & 	 0.982 	 & 	 0.760  \\
43	 & 	M 56	 & 	0.999	 & 	0.966	 & 	0.999	 & 	0.999	 & 	0.972	 & 	0.999	 & 	0.999	 & 	0.966 \\
47	 & 	M 75	 & 	0.346	 & 	0.474	 & 	0.862	 & 	0.346	 & 	0.999	 & 	0.523	 & 	0.648	 & 	0.474 \\
48	 & 	NGC 6934	 & 	0.795	 & 	0.622	 & 	0.915	 & 	0.795	 & 	0.999	 & 	0.897	 & 	0.622	 & 	0.990 \\
49	 & 	M 72	 & 	0.647	 & 	0.444	 & 	0.875	 & 	0.647	 & 	0.999	 & 	0.854	 & 	0.444	 & 	0.945 \\
50	 & 	NGC 7006	 & 	0.841	 & 	0.535	 & 	0.854	 & 	0.841	 & 	0.996	 & 	0.761	 & 	0.535	 & 	0.944 \\
51	 & 	M 15	 & 	0.425	 & 	0.278	 & 	0.425	 & 	0.548	 & 	0.806	 & 	0.956	 & 	0.278	 & 	0.903 \\
52	 & 	M 2	 & 	0.333	 & 	0.181	 & 	0.338	 & 	0.333	 & 	0.719	 & 	0.955	 & 	0.181	 & 	0.986 \\
54	 & 	Pal 13	 & 	0.366	 & 	0.155	 & 	0.772	 & 	0.366	 & 	0.986	 & 	0.711	 & 	0.155	 & 	0.982 \\
55	 & 	NGC 7492	 & 	0.318	 & 	0.717	 & 	0.318	 & 	0.929	 & 	0.717	 & 	0.760	 & 	0.962	 & 	0.788 \\
\hline
\multicolumn{10}{c}{Ultrafaint Dwarfs}\\
\hline
56	 & 	Segue 2	 & 	0.892	 & 	0.330	 & 	0.892	 & 	0.999	 & 	0.999	 & 	0.948	 & 	0.716	 & 	0.330 \\
57	 & 	UMa II	 & 	0.954	 & 	0.185	 & 	0.971	 & 	0.954	 & 	0.999	 & 	0.185	 & 	0.956	 & 	0.927 \\
59	 & 	Willman I	 & 	0.940	 & 	0.694	 & 	0.940	 & 	0.970	 & 	0.694	 & 	0.707	 & 	0.999	 & 	0.887 \\
60	 & 	Coma Ber	 & 	0.446	 & 	0.434	 & 	0.791	 & 	0.446	 & 	0.434	 & 	0.999	 & 	0.791	 & 	0.818 \\
62	 & 	Bootes I	 & 	0.493	 & 	0.774	 & 	0.999	 & 	0.493	 & 	0.878	 & 	0.999	 & 	0.774	 & 	0.983 \\
63	 & 	Segue 3\tablenotemark{b}	 & 	 0.945 	 & 	 0.833 	 & 	 0.945 	 & 	 0.968 	 & 	 0.985 	 & 	 0.999 	 & 	 0.833 	 & 	 0.938  \\
64	 & 	Pisces I	 & 	0.999	 & 	0.290	 & 	0.999	 & 	0.999	 & 	0.950	 & 	0.290	 & 	0.986	 & 	0.935 \\
\enddata
\tablenotetext{a}{Probability $P$ that a randomly selected artificial satellite would match the indicated segment of the Sgr stream as well as,
or better than, the indicated satellite.  Values $P<0.15$ are indicated in bold font.}
\tablenotetext{b}{No radial velocity information was available, so association statistics are calculated from angular position and distance alone.}
\label{results2.table}
\end{deluxetable*}

\begin{figure*}
\plotone{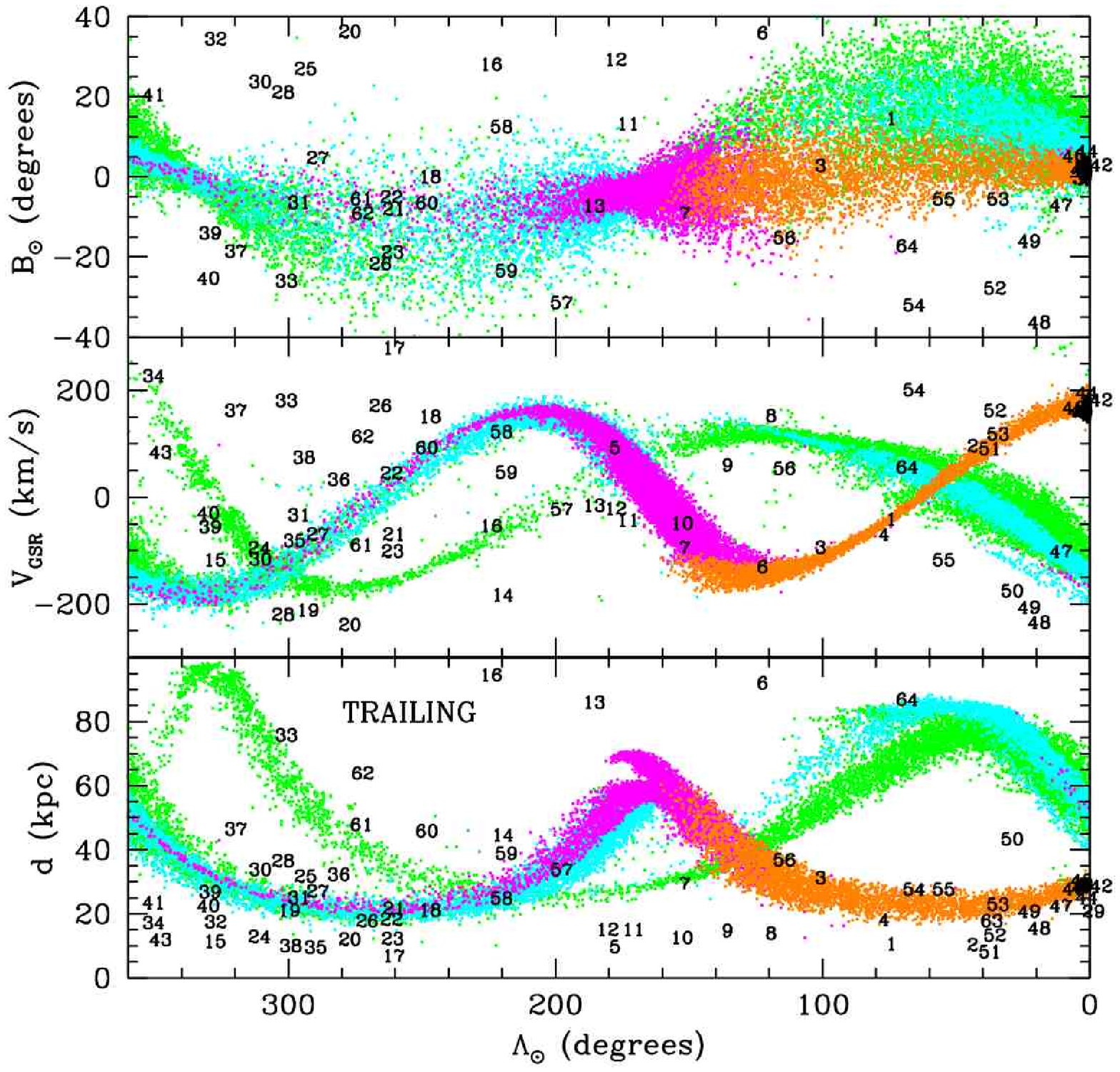}
\caption{Plot shows the predicted angular position $B_{\odot}$, distance $d$, and radial velocity $v_{\rm GSR}$ as a function of Sgr orbital longitude $\Lambda_{\odot}$
for the LM10 model of the Sgr trailing tidal stream.  Different colors for points represent debris
lost on different orbital passages (see \S \ref{model.sec}).  Milky Way stellar subsystems are overplotted, ID numbers correspond to those defined in
Table \ref{gcprops.table}.  Typically the stellar system is located at the bottom left corner of the corresponding ID label, although the locations of some ID numbers have been adjusted slightly to 
minimize confusion from overlapping labels.}
\label{TRAILdat.fig}
\end{figure*}

\begin{figure*}
\plotone{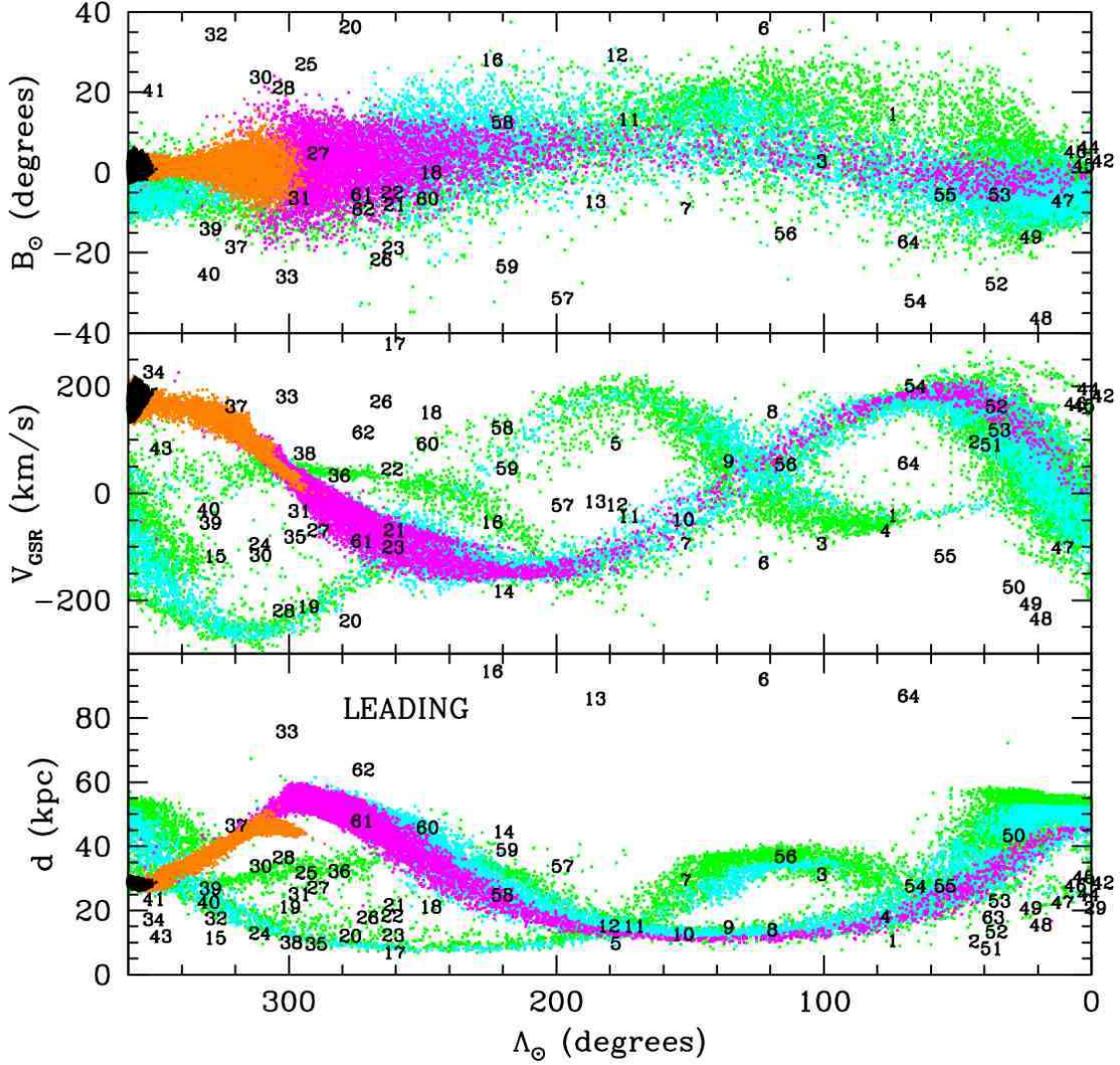}
\caption{As Figure \ref{TRAILdat.fig}, but for the  leading tidal stream.}
\label{LEADdat.fig}
\end{figure*}

\begin{figure*}
\plotone{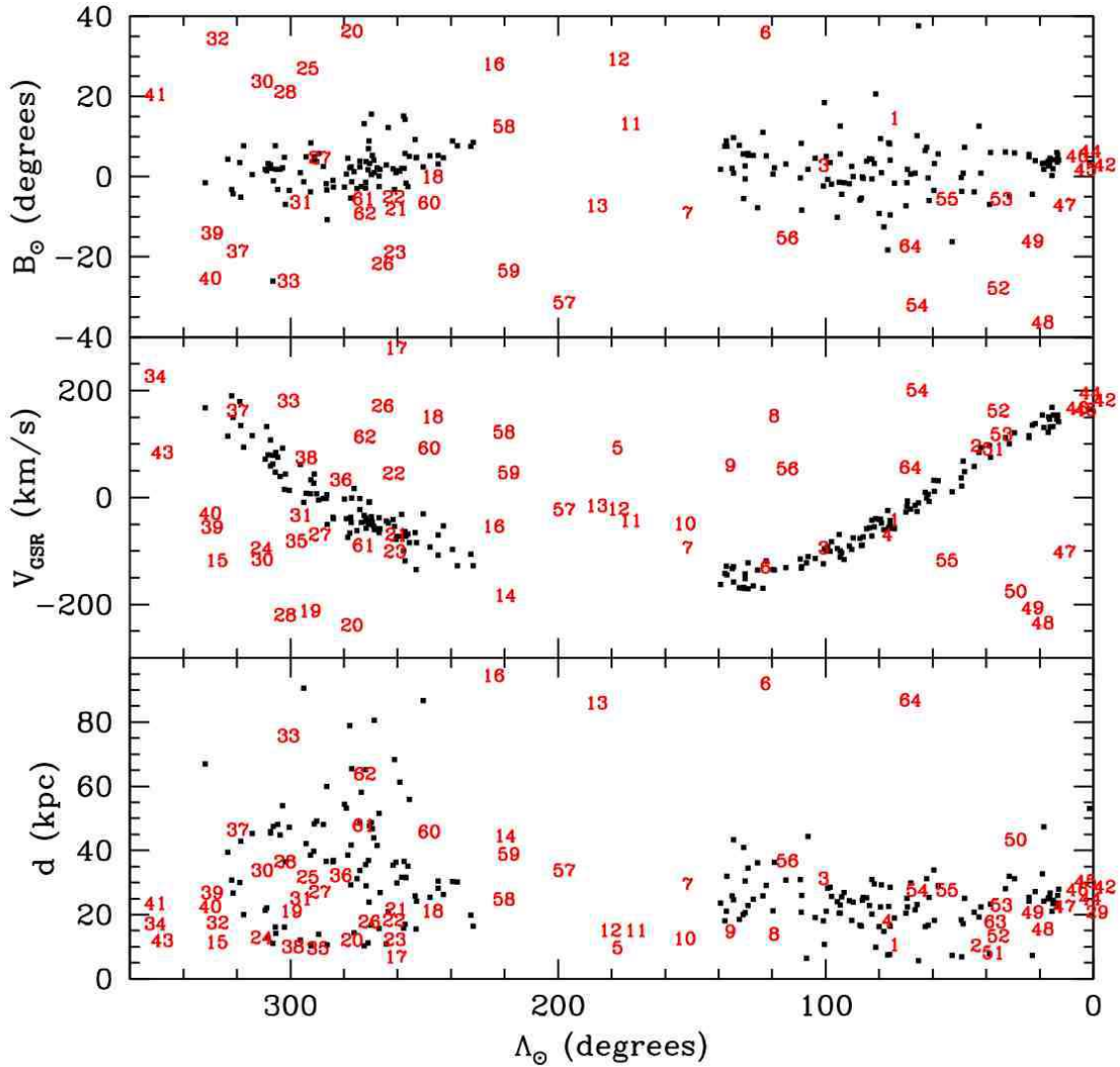}
\caption{As Figures \ref{TRAILdat.fig} and \ref{LEADdat.fig}, but black solid points indicate observed Sgr stream M-giants from the leading and trailing arm samples described by LM10.}
\label{MGdat.fig}
\end{figure*}

In Figures \ref{TRAILdat.fig} and \ref{LEADdat.fig} we overplot each of the satellites from Table \ref{gcprops.table} (coded by ID number) against the 
LM10 model of angular positions, distances, and radial velocities for the Sgr trailing/leading arm streams
respectively.
For comparison, in Figure \ref{MGdat.fig} we overplot the satellites against 2MASS M-giants whose radial velocities (Law et al. 2004; Majewski et al. 2004) are consistent
with membership in the Sgr stream (see Figure 1 of LM10).
Satellites that we consider are/are-not candidates 
for membership in the Sgr
stream based on the selection criteria $P_3 \leq 0.15$ or $P_7 \leq 0.15$
are summarized in 
Tables \ref{results1.table} and \ref{results2.table} respectively.\footnote{The four globular clusters AM 4, IC 4499, BH 176, and ESO 280-SC06, and the ultrafaint dwarf
Segue 3 do not presently have published radial velocities.  We therefore estimate $P_3$ and $P_7$ for these objects using a revised version of the $\chi$ statistic which is based upon
only distance and angular position.}
Since $P_3$ best corresponds to the observationally verified sections of the Sgr stream, 
we focus our discussion primarily on this statistic, except in those cases
when a better match is found using $P_7$.

\begin{figure*}
\plotone{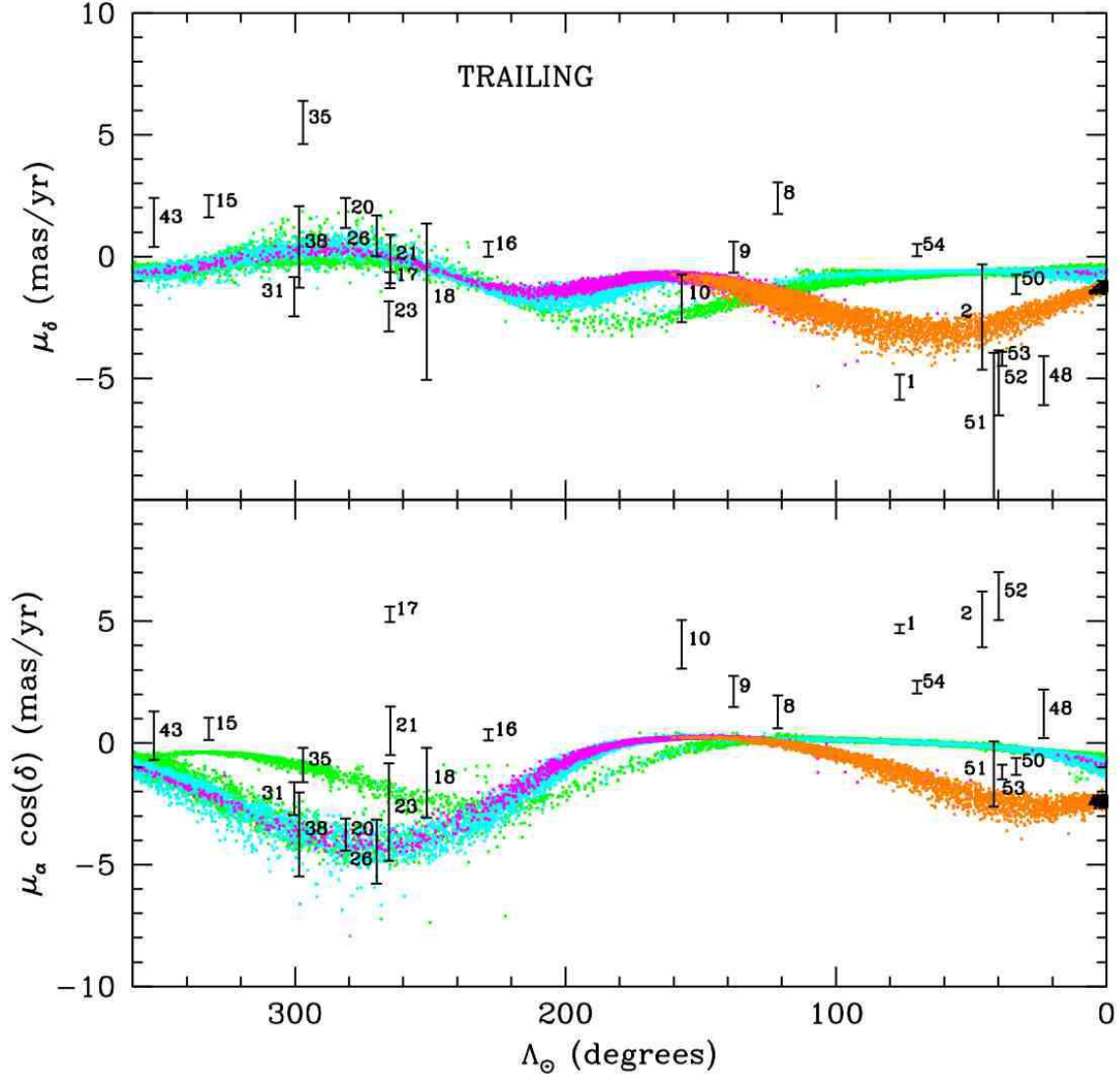}
\caption{Plots predicted proper motion  along right ascension ($\mu_{\alpha}$ cos$\delta$) and 
declination ($\mu_{\delta}$) as a function of Sgr orbital longitude $\Lambda_{\odot}$
for the LM10 model of the Sgr trailing tidal stream.  Different colors for points represent debris
lost on different orbital passages (see \S \ref{model.sec}).  Milky Way stellar subsystems are overplotted, ID numbers correspond to those defined in
Table \ref{gcprops.table}.  Errorbars on the proper motion measurements represent $1\sigma$.  The locations of ID numbers have been adjusted
to minimize confusion from overlapping labels.}
\label{TRAILpm.fig}
\end{figure*}
\begin{figure*}
\plotone{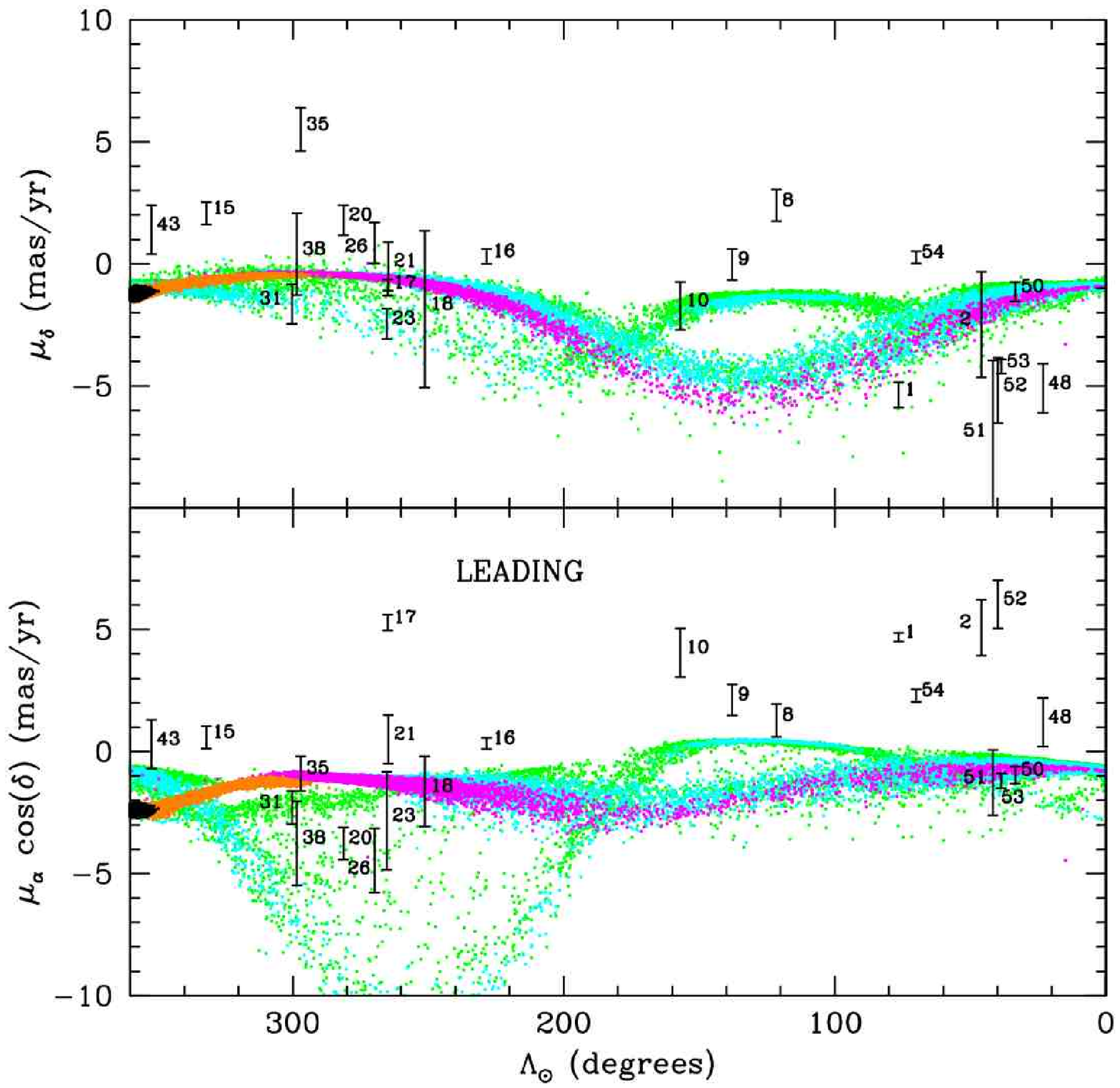}
\caption{As Figure \ref{TRAILpm.fig}, but for the leading tidal stream.}
\label{LEADpm.fig}
\end{figure*}

Given the large number of globular clusters
($7\pm2$)
that we expect to fall in the range $P \leq 0.15$ by virtue of chance superposition in angular position, distance, and radial
velocity with the Sgr stream, not all candidates listed in Table \ref{results1.table} will prove to be genuinely associated with Sgr.
Instead $P$ is a useful statistical indicator, suggesting that a system is sufficiently well-matched to warrant 
consideration; these objects are discussed further
in the subsections below.
In many cases, analysis of proper motions or other physical considerations can help confirm whether an association is
likely to be genuine.
Where available, proper motions for each of the satellites listed in Table \ref{gcprops.table} have been overplotted against those of the model Sgr stream
in Figures \ref{TRAILpm.fig} and \ref{LEADpm.fig} for trailing/leading arms respectively.
In the interests of brevity, we do not discuss satellites with both $P_3 > 0.15$ and $P_7 > 0.15$ (which are extremely unlikely to belong to Sgr) in detail
unless they exhibit some other point of interest, or have been described as Sgr cluster candidates previously in the literature.

In the following subsections we begin our discussion with the classical Sgr `core' globular clusters M 54, Arp 2, Terzan 7, and Terzan 8, 
and thereafter proceed to discuss individual systems in roughly alphanumerical order to aid the reader in easily locating particular sections of interest.
In most cases, we give the
$\Lambda_{\odot}$ coordinate and ID number  of each cluster in the corresponding subsection heading.
We group our discussion of the clusters M2, NGC 5466, NGC 5824, NGC 6426, and Rup 106
together in \S \ref{othergcs.sec}, and discuss the four clusters without radial velocity data (AM 4, BH 176, ESO 280-SC06, IC 4499)
in \S \ref{nodists.sec}.
Finally, we also discuss globular clusters that did not meet our initial $8 < r_{\rm GC} < 100$ kpc selection criterion briefly  in 
\S \ref{nearfargcs.sec}.
Our conclusions regarding the star cluster budget of Sgr are summarized in Section \ref{discussion.sec}.

\subsection{M 54, Arp 2, Terzan 7 and Terzan 8}

The four globular clusters M 54, Arp 2, Terzan 7, and Terzan 8 have traditionally  been associated with the Sgr dSph
(e.g., Ibata et al. 1994; Da Costa \& Armandroff 1995); these four `core' clusters lie within $\sim 1$ kpc of the
best-fitting Sgr debris plane and $10^{\circ}$ ($\sim 5$ kpc) of the Sgr dwarf core along the trailing tidal stream.
To within observational uncertainty, M 54 appears to be spatially coincident with the core of Sgr, and to lie at the same distance and radial velocity.
While M 54 is certainly associated with the Sgr dwarf (we calculate a less than $P_3 = 0.1$\% chance that an artificial globular cluster would chance to
match Sgr so well), its
velocity dispersion and luminosity profile  indicate that it formed independently of the Sgr core and later sank to its present position via dynamical friction
(Bellazzini et al. 2008, see also Majewski et al. 2003; Monaco et al. 2005; Siegel et al. 2007).

Arp 2 and Terzan 8 each have similarly strong probabilities for association.  
While proper motion information is not available for these clusters to confirm the association, the strength
of the match in angular position, distance, and radial velocity is sufficient to give us confidence in concurring with previous studies 
(see Table \ref{prevcands.table}) that M 54, Arp 2, and Terzan 8 are genuine Sgr members.
Indeed, out of a sample of 51 randomly chosen artificial globular clusters we should expect less than $0.05$
of them to match the Sgr stream so well by chance, let alone the 3 that are actually observed.

In contrast, Terzan 7 has a significantly lower confidence for association ($P_3 \approx 0.12$);
we expect $\sim 6$ satellites in a given sample of 51 to meet this threshold simply by chance.
However,  the high $P_3$ primarily reflects our adopted distance to Terzan 7 ($d = 23.2$ kpc), which is almost 5 kpc less than
the distance to the Sgr core ($D_{\rm Sgr} = 28$ kpc; Siegel et al. 2007) adopted in the LM10 model.  If the LM10 model had instead adopted
a slightly smaller distance for Sgr (reasonable estimates range from
$D_{\rm Sgr} = 24 - 28$ kpc; see Table 2 of Kunder \& Chaboyer 2009 for a summary), the quality of fit of Terzan 7 to the Sgr stream would improve considerably.  
Similarly, had we adopted a distance of $d = 25.75$ kpc for Terzan 7 (M. Siegel 2009, private communication)
based on new photometry akin to that used to set the Sgr distance scale in Siegel et al. (2007), we would have found $P_3 = 0.032$ and $P_7 = 0.052$.
While Terzan 7 is therefore not as robust a Sgr cluster candidate as M 54, Arp 2, or Terzan 8, physical association is still likely.
Future proper motion measurements should be able to verify this;
the proper motion of the Terzan 7 region of the Sgr stream is extremely well constrained by current observational data,
and if Terzan 7 is moving with the stream the LM10 model suggests that
we  should expect it to have a proper motion $\mu_{\alpha}$ cos$\delta = -2.4 \pm 0.1$ mas yr$^{-1}$ and $\mu_{\delta} = -1.4 \pm 0.1$ mas yr$^{-1}$.

\subsection{Berkeley 29 and Saurer 1 (ID=11/12, $\Lambda_{\odot}=177^{\circ}/182^{\circ}$)}
\label{Be29.sec}

Previous studies of the star cluster budget of the Sgr dwarf (e.g., Palma et al. 2002; Bellazzini et al. 2003a) have typically neglected sparse
open clusters located in the plane of the Galactic disk.  However, Carrarro \& Bensby (2009) argue that the 
metal-poor ($\feh = -0.44 \pm 0.18$ and $-0.38 \pm 0.14$ respectively; Carraro et al. 2004) cluster pair Be 29
and Sa 1 may be members of the Sgr stream.\footnote{Note that the declination 
given for Be 29, and the right ascension and declination
given for Sa 1 in Table 1 of Carraro \& Bensby (2009) are incorrect.}
Although typically regarded as open clusters, Be 29 and Sa 1 have ages, metallicities, and richness
intermediate between classical globular cluster and open cluster designations.  Since Be 29 in particular
(1) lies $\sim 2$ kpc above the Galactic plane, (2) is the most distant known open cluster (Tosi et al. 2004), and
(3) is brighter and richer than some globular clusters (e.g., AM 4, Whiting 1,  and Fornax-H1), we consider it reasonable that these two clusters
may indeed be part of the halo population rather than the disk.


While both clusters are approximately coincident with the Sgr L1 arm, Be 29 is the superior match with $P_3 = 0.027$ (compared to $P_3 = 0.190$ for Sa 1).
Visually, Figure \ref{LEADdat.fig} confirms this: Be 29 is superimposed on the L1 arm while Sa 1
lies off the projected path of the Sgr stream by $\sim 20^{\circ}$ ($4.1\sigma$) in Sgr orbital latitude $B_{\odot}$.
Both clusters lie at a distance $d = 13.2$ kpc consistent with that of the L1 arm, but 
Be 29 is a slightly better match to the L1 stream
in radial velocity (although still discrepant by $\sim 60$ \kms, or $3\sigma$) than is Sa 1 (off by $\sim 80$ \kms, or $4\sigma$).
The radial velocity of the L1 stream is not well constrained in this region however since it lies 
$\sim 50^{\circ}$ and $\sim 20^{\circ}$ respectively beyond the last 2MASS (see, e.g., Figure \ref{MGdat.fig}) and SDSS observations used to
constrain the LM10 model,
and it is therefore possible that the LM10 model at the location of Be 29 and Sa 1 may be in error by up to a few tens of kilometers per second.


The simplest hypothesis is that neither Sa 1 nor Be 29 are associated with the Sgr stream, and that both are unusual Galactic disk open clusters that happen to lie
near the nexus at which the L1 arm crosses through the disk and happen to have radial velocities resembling that of the stream.
Given the thousands of open clusters in the Galactic disk, it is perhaps unsurprising that some cluster might happen to have the requisite combination of properties.
This simple hypothesis is probably true for Sa 1 considering its low statistical probability for association with the Sgr stream 
($P_3 = 0.190$)\footnote{Alternatively it may instead be associated with some other tidal feature such as Monoceros/GASS; see Frinchaboy et al. 2006, Pe{\~n}arrubia  et al. 2005.} 
but the low $P_3 = 0.027$ for Be 29 may favor the Sgr-origin hypothesis.
Although a tentative measurement of the proper motion of Be 29 was presented by Dias et al. (2006), Frinchaboy (2006) demonstrated that the majority of stars used to make
this measurement were foreground field stars.  Only a single star had the correct magnitude and radial velocity to be a cluster member, and the proper motion of this star
($\mu_{\alpha}$ cos$\delta = -2.0 \pm 7.5$ mas yr$^{-1}$, $\mu_{\delta} = -5.7 \pm 7.5$ mas yr$^{-1}$; P. Frinchaboy 2010 [priv. communication]) is too uncertain to offer
a significant constraint on the relation between Be 29 and the Sgr stream.


If Be 29 is indeed a member of the Sgr stream, it is intriguing that this cluster 
is wrapped farther from the Sgr dwarf than all tidal debris conclusively observed to date,
and is the only known cluster (except possibly for Pal 12) associated with the leading tidal
arm.  The L1 arm at the location of Be 29 has been unbound from the Sgr core
for at least 2.5 Gyr, meaning that if Be 29 formed in the Sgr dwarf $\sim 4$ Gyr ago (Carraro et al. 2004 derive an age of 4.5 Gyr, while
Tosi et al. 2004 find $\sim 3.4 - 3.7$ Gyr) 
it must have been torn from Sgr shortly after its formation, and has since experienced at least 3 orbits around the Milky Way.
Given the similarities between Be 29, Whiting 1, and Terzan 7 in age, metallicity, luminosity, horizontal-branch type, and size (see discussion in \S \ref{discussion.sec})
we suggest that Be 29 may be a paradigm for the origins of some old, metal-poor `open clusters' in the Milky Way; namely, that they may be a class of sparse globular cluster
formed in dSph, stripped from their sheltering birth galaxies, and subjected to the gravitational tides of the Milky Way.


\subsection{IC 1257 (ID=39, $\Lambda_{\odot}=334^{\circ}$)}

IC 1257 has never previously been suggested as a possible Sgr cluster, but our analysis indicates that it resembles the L2 wrap at $\Lambda_{\odot} = 334^{\circ}$ to $P_7 = 0.112$.
This association is not particularly compelling however; while it closely matches the distance of the L2 wrap ($\sim 25$ kpc)
the angular position is inconsistent by $\sim 10^{\circ}$ and the radial velocity by $\sim 115$ \kms.   It is only due to the large apparent width
of the L2 stream in radial velocity ($\sigma_{\rm v, Sgr} \approx 60$ \kms) that the association statistic is calculated to be as good as $P_7 = 0.112$.  
Since $P_7 = 0.112$ corresponds to $\sim 6$ false associations out of a sample of 51 clusters, and
there is not even any concrete evidence that the L2 wrap actually exists in the Galactic halo, we 
doubt that IC 1257 is genuinely associated with the Sgr stream.

\subsection{M 3 (NGC 5272; ID=23, $\Lambda_{\odot}=265^{\circ}$)}

At a distance of 10.4 kpc, M 3 is too close to be associated with anything except the secondary L2 wrap
of the leading arm.  As illustrated in Figures \ref{LEADdat.fig} and \ref{MGdat.fig}, the radial velocity of M 3 is consistent with the L2 identification, but the angular position is
discrepant by $26.5^{\circ}$.
The importance of this angular discrepancy is mitigated somewhat by the large angular width of the L2 stream
($26.5^{\circ}$ corresponds to $3.8\sigma$), but there is a $P_7 = 12.2$\% chance (i.e., 6 out of a sample of 51) that a randomly chosen artificial cluster would appear to be associated
with some wrap of the Sgr stream more convincingly than is M 3.

Adopting the proper motions indicated in Table \ref{gcprops.table}, we observe (Figure \ref{LEADpm.fig}) that although the $\mu_{\delta}$  for M 3 is consistent with the L2 wrap
of the Sgr stream, $\mu_{\alpha}$ cos$\delta$ is inconsistent by $\sim 6.9$ mas yr$^{-1}$ (corresponding to $2.3\sigma$).
While at least four different values have been measured for the proper motion of M 3 (see compilation in Palma et al. 2002), none of these measurements agree more closely with 
the proper motion expected for the Sgr stream, and some (e.g., Cudworth \& Hansen 1993) are significantly poorer matches  to the estimated  $\mu_{\delta}$.
We therefore conclude that M 3 is probably not physically associated with the Sgr stream, although future constraints on the Sgr stream at extremely large separations from the dSph,
and measurements of the proper motion of M 3
to an accuracy $\sim$ 0.5 mas yr$^{-1}$ will help to check this conclusion.

\subsection{M 53 (NGC 5024; ID=21, $\Lambda_{\odot}=265^{\circ}$)}

M 53 is situated towards the North Galactic Cap and has been previously discussed as a possible 
candidate Sgr globular cluster by Palma et al. (2002; although these authors rule it out on the basis of its low metal content $\feh = -1.99$),
Bellazzini et al. (2003a), and Mart{\'{\i}}nez Delgado et al. (2004).
We find a mild probability for association between M 53 and the T1 ($P_{\rm 3} = 0.106$), L1 ($P_3 = 0.185$; barely failing the $P<0.15$ criterion),
and L2 ($P_{\rm 7} = 0.112$) arms.
Figures \ref{TRAILdat.fig} and \ref{LEADdat.fig} indicate that while M 53 matches the
T1 arm in angular position and distance it differs by 96 \kms (i.e., $4\sigma$) in radial velocity, and while it matches the L1/L2 arms in radial velocity it 
lies at the edge of both streams
and is simultaneously too close for the L1 arm and too distant for the L2 arm
(17.8 kpc versus $45\pm4$ kpc and $11\pm3$ kpc respectively).

While the proper motion (Figures \ref{TRAILpm.fig} and \ref{LEADpm.fig}) along the declination axis ($\mu_{\delta}$) 
is consistent with all three arms to within observational uncertainty, the proper motion
along right ascension ($\mu_{\alpha}$ cos$\delta$) is inconsistent by 4.7/1.7/10.2 mas yr$^{-1}$ ($4.3\sigma/1.7\sigma/4.1\sigma$) 
with the T1/L1/L2 arms respectively.
Given these difficulties in associating M 53 conclusively with any particular arm of the Sgr stream and the likelihood that $\sim 5$ clusters 
with similarly good $P$ statistics will be found in our sample of 51 clusters by chance,
we  conclude that M 53 is probably not physically associated with the Sgr system.

\subsection{NGC 288 (ID=1, $\Lambda_{\odot}=76^{\circ}$)}

NGC 288 lies within $\sim 1$ kpc of the Sgr plane in the direction of the South Galactic Pole, and (as pointed out by Bellazzini et al. 2003a)
has a radial velocity very close to that expected for the T1 trailing arm.
Although the distance to NGC 288 ($d = 8.8$ kpc) is consistent with that of the M-giants (Figure \ref{MGdat.fig}), this is likely due to the large observational
uncertainty in the photometric parallaxes of the M-giants; NGC 288 lies $\sim 10$ kpc closer than the LM10 model of the T1 stream.
Since the association probability $P_3 = 0.095$ 
corresponds to $\sim$ 5 false positives anticipated in our sample of 51 globular clusters,
we are inclined to doubt the association of NGC 288 with the Sgr stream.  Table \ref{results1.table} indicates that 
significant matches to other segments of the Sgr stream are even less likely.
This negative hypothesis is borne out by the proper motions: Figure \ref{TRAILpm.fig} demonstrates 
that the $\mu_{\alpha}$ cos$\delta$ and $\mu_{\delta}$ of NGC 288 are discrepant with the T1 stream
at this location by  5.9 mas yr$^{-1}$ and 2.4 mas yr$^{-1}$ respectively.  Combining the uncertainty in the observed proper motions and the
variance in proper motions of particles in the LM10 model,
this corresponds to an offset of $\sim 22\sigma$.
We are therefore confident in our conclusion that, despite its proximity to the Sgr plane, NGC 288 is not physically associated with the Sgr stream.


 \subsection{NGC 2419 (ID=13, $\Lambda_{\odot}=190^{\circ}$)}
 
Irwin (1999) and Newberg et al. (2003) note that the distant ($d = 84$ kpc) Galactic globular cluster NGC 2419 ($[l,b] = [180^{\circ}, 25^{\circ}]$)
lies near the Sgr debris plane, and suggest that it may be associated with the Sgr trailing tidal arm.
Indeed, Figure \ref{TRAILdat.fig} shows that NGC 2419 lies in projection against the T1 arm with $B_{\odot} \approx -9^{\circ}$.
However, the large distance to the cluster is poorly matched by the LM10 model, which predicts that the T1 stream
reaches apogalacticon at a distance $\sim 61$ kpc around at $\Lambda_{\odot} \sim 170^{\circ}$, and approaches closer to the Milky Way again at greater orbital longitudes.
NGC 2419 is formally inconsistent with the LM10 model by $\sim 38$ kpc in distance and $\sim 160$ \kms in radial velocity, giving a poor association statistic
$P_3 = 0.386$.

Direct examination of the T1 M-giant stream observed by Majewski et al. (2003; their Figure 10) is less clear; while it admits the possibility that the observed stream
may have a distance $\sim 80$ kpc at $\Lambda_{\odot} = 190^{\circ}$ (perhaps tracing a previously larger orbit for Sgr, which has since sunk deeper into the Galactic
gravitational potential via dynamical friction), the stream at this location is difficulty to trace because it is poorly populated by M-giants and has an uncertain distance calibration.
While the LM10 model of the T1 arm is not directly constrained at $\Lambda_{\odot} = 190^{\circ}$, it agrees well with observations of this arm at both
$\Lambda_{\odot} \sim 160^{\circ}$ (i.e., the M-giants) and $\Lambda_{\odot} \sim 225^{\circ}$ (see discussion in \S \ref{segueI.sec}), suggesting that it should not be wildly
inaccurate in the longitude range of NGC 2419.  We therefore conclude that NGC 2419 is unlikely to be genuinely associated with the Sgr dSph.

\subsection{NGC 4147 (ID=18, $\Lambda_{\odot}=251^{\circ}$)}


NGC 4147 (1.4 kpc from the Sgr midplane) 
has been previously discussed as a potential member of the Sgr system 
by Bellazzini et al. (2003ab) and Palma et al. (2002).
This cluster is a formally fair match to the T1 principle wrap of the trailing tidal arm (as observed in the direction of the North Galactic Cap),
with $P_3 = 0.058$.
As indicated by Figure \ref{TRAILdat.fig} however, NGC 4147 lies near the edge of the T1 stream 
in projection and is discrepant from its radial
velocity by $\sim 86$ \kms.  NGC 4147 is therefore not an extremely attractive candidate for membership in the Sgr stream
because $\sim 3$ such clusters in our sample of 51 are expected to appear at least as closely associated by chance.

While proper motions have been  measured for NGC 4147 (Brosche et al. 1991), they do not conclusively confirm or deny membership in the Sgr stream;
as illustrated in Figure \ref{TRAILpm.fig}
$\mu_{\alpha}$ cos$\delta$ is discrepant from the T1 stream by 2.3 mas yr$^{-1}$ (corresponding to $\sim 1.5\sigma$),
while $\mu_{\delta}$ is consistent with the T1 stream.  However,  the uncertainties on the observed proper motions (particularly for $\mu_{\delta}$)
are so large that they contain little discriminatory power.
We are therefore unable to determine conclusively the whether NGC 4147 is a member of the Sgr stream, although it appears to be only a weak candidate.


\subsection{NGC 5053 (ID=22, $\Lambda_{\odot}=266^{\circ}$)}

NGC 5053 lies near M 53 in the North Galactic Cap (separation $\sim 1^{\circ}$) and at a similar distance
(as discussed by Palma et al. 2002; Bellazzini et al. 2003a), although
the two clusters have radial velocities that differ by $\sim 125$ \kms.  As illustrated by Figure \ref{TRAILdat.fig}, the radial velocity of
NGC 5053 ($v_{\rm GSR} = 34.1 \pm 0.4$ \kms) is a much better match to that expected for the T1 stream ($v_{\rm GSR} = 25 \pm 17$ \kms) than was M 53.
This results in a much improved value $P_3 = 0.04$;  only $\sim 2$ such systems are expected to occur by chance in our sample of globular clusters.

Lauchner et al. (2006; see also Chun et al. 2010) have reported a tentative detection of tidal streams
emanating from the cluster, stretching from $(\alpha,\delta) = (199^{\circ}, 18^{\circ})$ towards $(195^{\circ}, 15^{\circ})$.
If genuine, these streams may weaken the case for association of NGC 5053 with the Sgr stream since they suggest a
motion oriented at a transverse angle $\sim 60^{\circ}$
to the Sgr plane.  However, these streams may be too close to the cluster ($\sim 3-4$ tidal radii; Lauchner et al. 2006) to provide a good indication of the actual
direction of the cluster's orbit (see, e.g., Johnston et al. 2002; Montuori et al. 2007; Klimentowski et al. 2009; Odenkirchen et al. 2009), and cannot yet be used to
eliminate NGC 5053 from the Sgr family.

Unfortunately, no proper motion has been measured for NGC 5053, and it is therefore not possible to conclusively determine its membership status in the Sgr stream.
If it proves to be a genuine member, we might expect the proper motion to be similar to
$\mu_{\alpha}$ cos$\delta = -4.2 \pm 0.5$ mas yr$^{-1}$ and $\mu_{\delta} = -0.1 \pm 0.1$ mas yr$^{-1}$.
At present however, the case for membership appears to be fairly good.  If confirmed, NGC 5053 would be the most metal poor of the Sgr globular clusters
with $\feh = -1.98$ (Marin-Franch et al. 2009).

\subsection{NGC 5634 (ID=27, $\Lambda_{\odot}=293^{\circ}$)}

NGC 5634 lies within 0.5 kpc of the Sgr plane, and is consistent with the T1 segment of the trailing arm stream which has been wrapped around
the Milky Way into the direction of the North Galactic Cap.  As indicated by Table \ref{results1.table}, 
there is a probability $P_3 = 0.4$\% for a randomly chosen halo globular cluster to match the Sgr stream this well, corresponding to $\sim 0.2$ such clusters in a given sample of 51.
As illustrated in Figure \ref{TRAILdat.fig},
NGC 5634 matches the model T1 stream to within $\sim 6^{\circ}$ ($1\sigma$) in $B_{\odot}$, 2 kpc ($< 1\sigma$) in distance, and 4 \kms ($< 1\sigma$) in $v_{\rm GSR}$.
The extremely low metallicity 
of NGC 5634 ($\feh = -1.94$; Bellazzini et al. 2002) is consistent
with that of  the three Sgr core clusters M 54, Arp 2, and Terzan 8, which have $-1.2 \gtrsim \feh \gtrsim -1.8$.
We agree with the possibility suggested by Bellazzini et al. (2002; 2003a) that NGC 5634 
is an extremely strong candidate for a Sgr globular cluster.

Proper motion measurements  have not yet been obtained to test this hypothesis, but if NGC 5634 is indeed associated with the T1 stream we might expect it to have
 $\mu_{\alpha}$ cos$\delta \approx -3.8 \pm 0.4$ mas yr$^{-1}$ and $\mu_{\delta} \approx +0.2 \pm 0.1$ mas yr$^{-1}$.

\subsection{Pal 2 (ID=7, $\Lambda_{\odot}=154^{\circ}$)}

Pal 2 is a heavily obscured ($E(B-V) = 1.24\pm0.07$; Harris et al. 1997) globular cluster that lies in the direction of the Galactic anticenter.
The possibility that Pal 2 may be part of the T1 trailing arm of Sgr has been previously noted by Lynden-Bell \& Lynden-Bell (1995), Irwin (1999), and Majewski et al. (2004). 

As illustrated in Figure \ref{TRAILdat.fig}, Pal 2 lies in the T1 trailing arm stream in projection, and has a radial velocity consistent with this identification, but
at a distance $d = 28\pm4$ kpc (Harris et al. 1997), roughly half that expected for the T1 stream at this orbital longitude in the LM10 model ($d = 55\pm 4$ kpc).
This 27 kpc distance mismatch translates to a $P_3 \approx 12$\% chance that a random halo globular cluster would match the Sgr stream as well as, or better than, Pal 2
by chance.  Since we expect six such false positives in our sample of 51 clusters, we are inclined to treat the potential association of Pal 2 with the Sgr stream with
skepticism.

It is possible  since Pal 2 is so heavily obscured that its distance (Harris et al. 1997) may have been underestimated, although
we note that Sarajedini et al. (2007) derive a similar distance $d = 27\pm2$ kpc based on main-sequence fitting.
Alternatively, the LM10 model may systematically overestimate the distance
to the Sgr stream at this location.  As discussed by Majewski et al. (2004), Pal 2 appears to lie only a few kpc closer than the closest of the M giants observed in this
region (Majewski et al. 2003).  However, the M-giants distances in this region may be unreliable given the evolving metallicity distribution function (MDF) along the Sgr
tidal streams (Chou et al. 2007, 2010); the LM10 model was designed to account for the evolution in the MDF
and matches the observed M-giants in apparent magnitude (see Figure 18 of LM10).
Unfortunately, the proper motion of Pal 2 has not been measured and it is therefore not presently possible to determine conclusively the relation of Pal 2 to the Sgr stream.

 \subsection{Pal 5 (ID=31, $\Lambda_{\odot}=300^{\circ}$)}

Pal 5 lies near NGC 5634, and within $\sim 4$ kpc of the Sgr plane.
At a distance of 23.2 kpc it falls inside the loop of Sgr L1 leading arm debris, and is consistent with the distance expected for the T1 stream wrapped
into the northern Galactic hemisphere (see Figure \ref{TRAILdat.fig}).
However, the
radial velocity of Pal 5 is inconsistent with this identification by $\sim 83$ \kms ($3.8\sigma$).
This radial velocity mismatch gives a probability statistic $P_3 = 0.045$, corresponding to $\sim 2$ such false positives expected in our sample of 51 globular clusters.

The situation of Pal 5's possible association with Sgr
has been discussed by Bellazzini et al. (2003a) based on its close proximity to the Sgr plane, by Palma et al. (2002) on the basis of the location of
its orbital pole,
and with regard to its proper motion by Scholz et al. (1998).
It has also been noted (Palma et al. 2002) that Pal 5 has a similar mass and concentration to the candidate Sgr globular clusters Arp 2, Terzan 7, Terzan 8, and Pal 12,
and that its alpha element abundance is similar to that observed in M 54, Ter 7, and Pal 12 (Sbordone et al. 2005).  While these latter authors suggest that Pal 5 may be
a member of the Sgr stream based on its chemical properties, they note that its orbital characteristics suggest it is more likely a candidate to be associated with older rather
than younger tidal debris phases.
Indeed, Palma et al. (2002) demonstrate that while the orbital poles 
of Sgr and Pal 5 are somewhat similar (separated by about $27^{\circ}$),
the energy and momentum are sufficiently different
to indicate that Sgr and Pal 5 have rather different orbits.  
Similarly, Figure \ref{TRAILpm.fig} indicates that the proper motion of Pal 5 is inconsistent with that of the T1 stream
by 1.3/1.9 mas yr$^{-1}$ ($1.7\sigma/2.3\sigma$) for $\mu_{\alpha}$ cos$\delta$
and $\mu_{\delta}$ respectively. 



Pal 5 has also been shown
to have lengthy ($\sim 20^{\circ}$), well-defined tidal tails of its own (Odenkirchen et al. 2001, 2003; Rockosi et al. 2002; Grillmair \& Dionatos 2006) 
and these provide another indicator of the direction of motion (although see Odenkirchen et al. 2009).
A comparison of the direction of these tidal tails (see, e.g., Fig. 11
of Rockosi et al. 2002) to the direction of the
orbital path of Sgr on the sky
shows them to be nearly perpendicular to one another (i.e., roughly along $B_{\odot}$ at constant $\Lambda_{\odot}$).
We therefore conclude that Pal 5 is unlikely to be a member of the Sgr globular cluster system.

\subsection{Pal 12 (ID=53, $\Lambda_{\odot}=39^{\circ}$)}

Pal 12 has been the subject of vigorous debate in recent years, and claims of possible association with the Sgr streams
have been made by Mart{\'{\i}}nez-Delgado et al. (2002), Palma et al. (2002), Bellazzini et al. (2003ab), Majewski et al. (2004),
and on the basis of full space velocities by Dinescu et al. (2000).
The chemical abundance of Pal 12 ($\feh = -0.83$; Marin-Franch et al. 2009, see also Cohen et al. 2004,
Armandroff \& Da Costa 1991) and low \afeh $\,$ signatures characteristic of the Sgr family have also led Cohen et al. (2004)
and Sbordone et al. (2005) to conclude that Pal 12 is likely to be associated with the Sgr stream.

We find that Pal 12 presents an unusual case, in that it has a relatively 
convincing probability of association with both the L1 ($P_{\rm 3, L1Y} = 0.048$) and T1 ($P_{\rm 3, T1Y} = 0.012$) arms.
As illustrated in Figures \ref{TRAILdat.fig} and \ref{LEADdat.fig}, this curious circumstance arises because Pal 12 lies in a region of the South Galactic Cap at which
the T1 arm overlaps in angular position, distance,
and radial velocity with the L1 arm which
has passed through the northern Galactic hemisphere, through the Galactic disk, and is completing its wrap through  the southern Galactic hemisphere.
The angular position of Pal 12 matches the L1 arm best (Figure \ref{LEADdat.fig}), 
lying only $5.5^{\circ}$ ($\sim 1.4\sigma$) from the centroid of the stream, compared to $9^{\circ}$ ($\sim 2.6\sigma$) from the centroid 
of the T1 arm (Figures \ref{TRAILdat.fig} and \ref{MGdat.fig}).  
The distance of Pal 12 ($19.1$ kpc) best matches the T1 arm though, lying 4 kpc ($\sim 1.4\sigma$) closer than the T1 arm and 11 kpc ($\sim 3\sigma$) closer than the L1 arm.
The radial velocity is approximately equally matched to both segments of the Sgr stream, matching the L1 arm to within 44 \kms 
($\sim 1.3\sigma$) and the T1 arm to within 16 \kms ($\sim 1.5\sigma$).

This situation is not clarified by the observed proper motions:  As illustrated in Figures \ref{TRAILpm.fig} and \ref{LEADpm.fig}, 
Pal 12 is discrepant from the T1 arm by 1.3/1.5 mas yr$^{-1}$
(3$\sigma$/3.5$\sigma$) in $\mu_{\alpha}$ cos$\delta$/$\mu_{\delta}$ respectively, and is discrepant from
the L1 arm by 0.5/2.7 mas yr$^{-1}$ (1.4$\sigma$/6.2$\sigma$).
The uncertainties given on the proper motion of Pal 12 are particularly small (0.3 mas yr$^{-1}$) however; if these have been underestimated the significance
of the proper motion discrepancies decreases commensurately.


If one supposes that Pal 12 is associated with the T1 arm (as favored by the distance and proper motion estimates), it must be explained
why the angular position of Pal 12 places it at the extreme outer edge of the trailing stream and why the proper motions deviate from the model
by a total of $\sim 4\sigma$.  In contrast, if one supposes that Pal 12 is associated with the L1 arm (as favored by the angular coordinates,
as well as by Dinescu et al. 2000  and Majewski et al. 2004) it must somehow be explained why Pal 12 lies at a smaller
distance than predicted for the L1 stream (20 kpc versus 30 kpc) and why the proper motions are inconsistent by over $6\sigma$.
Since the LM10 model of the leading tidal stream wrapped so far (nearly $360^{\circ}$) from the Sgr dwarf is significantly less well constrained than the trailing arm
stream in the same region of the sky (see, e.g., Majewski et al. 2003), it is perhaps more plausible that Pal 12 is associated with the L1 arm.
 

Given the low probability ($P_3 = 0.012$, corresponding to $\sim 0.6$ clusters out of a randomly drawn sample of 51) 
for a globular cluster to appear to  match the Sgr stream 
as well as Pal 12 by chance, in combination with the unique chemical fingerprint characteristic of the Sgr globular clusters,
it seems likely that Pal 12 is physically associated with the Sgr stream.
In the future, an independent proper motion study of Pal 12 will hopefully 
check which arm of the Sgr stream (if either) Pal 12 is associated with.
 
 \subsection{Pal 13 (ID=54, $\Lambda_{\odot}=70^{\circ}$)}

Pal 13, located towards the South Galactic Cap, has a distance
and radial velocity consistent with the L1 wrap of the leading tidal stream (see Figure \ref{LEADdat.fig}).  At 15.1 kpc from the Sgr midplane however 
(corresponding to $B_{\odot} = -33.5^{\circ}$) it is a poor match to the predicted location of L1 tidal debris.  The L1Y arm 
has a mean value of $B_{\odot} = +0.3^{\circ}$ with an angular width
 $\sigma_{B, {\rm Sgr}} = 3.2^{\circ}$, placing Pal 13 over $10\sigma$ away from the stream and giving a $P_3 = 37$\% probability that a randomly chosen
 artificial globular cluster would match the Sgr stream better than Pal 13.
If the older ($> 3$ Gyr old) debris from the LM10 is included in the analysis, the angular width of the stream increases slightly, placing Pal 13 only
$6.7\sigma$ away from the L1 stream ($P_{\rm 7, L1} = 0.155$), but neither value is compelling evidence for association.

Strong evidence against the association of Pal 13 with the Sgr stream is provided by the proper motions
measured by Siegel et al. (2001; Figure \ref{LEADpm.fig}), which are inconsistent
with the  $\mu_{\alpha}$ cos$\delta$ and $\mu_{\delta}$ of the L1 arm by 3.2/3.1 mas yr$^{-1}$ ($7.3\sigma/4.8\sigma$) respectively.
In contrast to Bellazzini et al. (2003a), we therefore concur with Palma et al. (2002) that Pal 13 is not associated
with the Sgr dwarf.

\subsection{Pal 15 (ID=37, $\Lambda_{\odot}=324^{\circ}$)}

Pal 15 lies on the opposite side of the Galactic Center towards $(l,b) = (20^{\circ}, 25^{\circ})$ at a distance of 44.6 kpc.  As indicated by Table \ref{results1.table}
it has a formal probability $P_3 = 0.128$ for association with the L1 arm of the Sgr tidal stream.
This association is relatively weak given the corresponding expectation that $\sim 7$ globular clusters of our sample of 51 would match the Sgr stream better than Pal 15 by
chance.  While Pal 15 is a reasonable match to both the distance and radial velocity of the L1 arm at $\Lambda_{\odot} = 324^{\circ}$, it lies $\sim 20^{\circ}$ ($\sim 16$ kpc) away from the bulk
of the L1 tidal debris at this orbital longitude.  Although minor discrepancies in angular position might be rationalized in regions where the LM10 model of the Sgr streams is poorly constrained
(e.g., the L2/T2 arms), Pal 15 lies within $35^{\circ}$ of the Sgr core and its angular position disagrees strongly with observations of the leading arm as traced in 2MASS (e.g., Figure 3 of Majewski
et al. 2003).  The strong 2MASS M-giant constraints on the width of the Sgr debris stream at this orbital longitude rule out Pal 15 as a Sgr globular cluster.

\subsection{Whiting 1 (ID=3, $\Lambda_{\odot}=103^{\circ}$)}
\label{whiting1.sec}


Whiting 1 (Wh 1) is a relatively poorly known, recently discovered (Whiting et al. 2002) globular cluster 
that is fairly young ($\sim 6.5$ Gyr) and metal rich ($\feh = -0.65$; Carraro 2005).
On the basis of a distance estimate $d = 29.4$ kpc and radial velocity measurement $v_{\rm GSR} = -105 \pm 1.8$ \kms,
Carraro et al. (2007) suggested that it may be associated with the trailing arm of the Sgr stream in the direction of the South Galactic Cap.
We confirm this association, finding that Whiting 1 lies within $\sim$ 0.2 kpc of the Sgr plane (closer than any other cluster save M 54),
and that its angular position, heliocentric distance, and radial velocity match those of the T1 stream to extremely high accuracy (see Figs. \ref{TRAILdat.fig} and \ref{MGdat.fig}).
We calculate a probability $P_3 \sim 0.1$\% (corresponding to 0.05 clusters out of a sample of 51) that a globular cluster 
would match the Sgr stream as well as Whiting 1
by chance.  While proper motions for this cluster are not available to confirm the hypothesis, 
it appears that Whiting 1 can be associated with the Sgr system with a similar confidence as for
M 54, Arp 2, and Terzan 8.

As indicated by Table \ref{results1.table}, there is also a lower confidence association of Whiting 1 with the secondary wrap of the Sgr leading arm (L2).
However, since this association is at much lower confidence ($P = 6.4$\%, corresponding to an expectation of
$\sim 3$ clusters by chance in our sample of 51) than the T1 association
(and indeed it is not certain if L2 debris is even present in the Galactic halo) we therefore believe that the apparent L2 association is
not significant.

\subsection{M 2, NGC 5466, NGC 5824, NGC 6426, Rup 106}
\label{othergcs.sec}

Bellazzini et al. (2003a) discuss the possible association of these globular clusters with the Sgr stream,
and Sbordone et al. (2005) note the chemical similarity between Rup 106 and the Sgr family.
However, none of these clusters present significant evidence for association with any wrap of the Sgr arms,
with probabilities of $P_3 = 33$\%, 53\%, 29\%, 80\%, and 62\% that a randomly chosen globular cluster in the Galactic halo would by chance appear
to match the Sgr stream better than M 2, NGC 5466, NGC 5824, NGC 6426, or Rup 106 respectively.
While the poor match of the angular positions, distances,
and radial velocities of these clusters is sufficient to exclude them as Sgr members, the proper motion reinforces this
conclusion for M 2.

\subsection{AM 4, BH 176, ESO 280-SC06, IC4499}
\label{nodists.sec}

Radial velocities have not been measured for these four clusters, and their association statistics have therefore been estimated by repeating the analysis of Section
\ref{quantifying.sec} using distance and angular position alone.
These statistics $P_3 > 0.36$ and $P_7 > 0.45$ for all four clusters, indicating that neither are likely to be associated with the Sgr stream.
As illustrated in Figures \ref{TRAILdat.fig} and \ref{LEADdat.fig} 
(see ID numbers 25, 29, 32, and 41), ESO 280-SC06 is barely consistent with the angular position of the T2 stream, but lies
much too close to the sun (21.7 kpc vs. 79.2 kpc).  Neither BH 176 nor IC 4499 lie within $\sim 25^{\circ}$ of any simulated stream debris.
Carraro (2009) make the case that AM 4 might be associated with Sgr based on their
newly derived distance $d = 33^{+3}_{-4}$ kpc.  However, the closest angular match to AM 4 is provided by the L1 arm (whose centroid lies $24.6^{\circ}$ away,
or $3.7\sigma$), and both the value $d = 29.9$ kpc quoted in Table \ref{gcprops.table} and the revised estimate of Carraro (2009)
are strongly inconsistent with the distance of 53 kpc expected for the L1 arm at this angular position.
While radial velocities for these four clusters would verify our analysis, we strongly doubt that any of them is associated with the Sgr stream.

\subsection{Globular Clusters at Other Galactocentric Radii}
\label{nearfargcs.sec}

As described in \S \ref{gcsel.sec} above, we have focused our attention on globular clusters in the range of Galactocentric radii
$8 < r_{\rm GC} < 100$ kpc as a generous interval surrounding the predicted locations of Sgr tidal debris in the LM10 model.
However, it is conceivable that clusters currently located at other radii may also have originated in the Sgr dwarf and have been missed
in our analysis due to the effects of dynamical friction or other limitations
of the LM10 model.

While dynamical friction is not expected to substantially alter the path of individual globular clusters in the halo of the Milky Way
(the inspiral time for a cluster at a radius of 8 kpc in the Galactic potential is expected to be much longer than a Hubble time; see, e.g., Eqn 8-24
of Binney \& Tremaine [2008, p. 655]), it is expected to have an appreciable effect on the path of the Sgr dSph itself which was not incorporated in the LM10 model.
Since Sgr  should have sunk deeper into the gravitational potential of the Milky Way over its lifetime, tidal debris (and any corresponding globular clusters) torn
from Sgr at early times in its interaction might be expected to lie at systematically larger distances than more recent epochs of debris.
Only two globular clusters are found in the Harris (1996) catalog at distances $r_{\rm GC} > 100$ kpc; E 1 and Pal 4 at distances of $r_{\rm GC} = 123$ and 112 kpc respectively.
While E1 lies $\sim 100$ kpc from the Sgr plane, Pal 4 is only $\sim 15$ kpc away from this plane.  At $\Lambda_{\odot} = 239^{\circ}$, it is perhaps plausible that Pal 4 may
be linked to the Sgr stream in a similar manner to NGC 2419 (see also Palma et al. 2002).  
If the T1 arm continues to rise in distance for $\Lambda_{\odot} \gtrsim 160^{\circ}$ instead of turning around
as seen in the LM10 model, this arm may pass through both NGC 2419 and Pal 4.
This picture is perhaps unlikely, however, given the recent confirmation of the T1 stream at $\Lambda_{\odot} \sim 225^{\circ}$ with properties similar to those of the LM10
model predictions (see discussion in \S \ref{segueI.sec}).

Since the Milky Way gravitational potential adopted by LM10 is imperfect, it is also possible that some Sgr tidal debris may fall inside $r_{\rm GC} = 8$ kpc.
Indeed, three globular clusters at such radii (M 5, NGC 6356, and Terzan 3) have been discussed by previous authors (Palma et al. 2002; Bellazzini et al. 2003a)
as potential Sgr cluster candidates based either on their near-alignment with the Sgr plane or on their similarity of orbital pole
families.  NGC 6356 in particular lies on the edge of our selection annulus ($r_{\rm GC} = 7.6$ kpc) at angular coordinates
$(\Lambda_{\odot}, B_{\odot} = (336.6^{\circ}, -5.2^{\circ}))$.  With a heliocentric distance of $d = 15.2$ kpc it is most closely matched by
the L2 wrap of the leading arm, which is expected to lie at $d \approx 25$ kpc.  However, the radial velocity of NGC 6356 is $v_{\rm GSR} = +64$ \kms,
which is inconsistent with the LM10 prediction of the L2 arm by more than 250 \kms.
If we redo our construction of a statistical comparison sample  from \S \ref{gcsel.sec} - \ref{quantifying.sec} to incorporate globular clusters at $r_{\rm GC} < 8$ kpc,
we find that none of the clusters in this radius range have association statistics better than $P_3$ or $P_7 \sim 30$\%.  To within the limitations of the LM10 model,
we are therefore confident in our decision
to exclude these clusters from detailed consideration.

\section{ANALYSIS OF INDIVIDUAL ULTRAFAINT SYSTEMS}
\label{compareUF.sec}

In Figures \ref{TRAILdat.fig} and \ref{LEADdat.fig} we overplot each of the nine ultrafaint Milky Way satellites from Table \ref{gcprops.table} against the 
LM10 model of angular positions, distances, and radial velocities for the Sgr trailing/leading arm streams
respectively.
As indicated by Figures \ref{RandomUF.fig} and \ref{nofpuf.fig}, the ultrafaint satellites are more likely to appear to match the Sgr stream by chance than the globular cluster sample
by virtue of the footprint of the sky area that has been surveyed by the SDSS.
Although Figure \ref{nofpuf.fig} suggests that there is no population of these satellites obviously associated with the Sgr stream,
for consistency with the globular cluster analysis we have listed satellites with $P \leq 0.15$
in Table \ref{results1.table} and $P > 0.15$ in Table \ref{results2.table}, and discuss all nine satellites individually below.

\subsection{Bo\"{o}tes I (ID=62, $\Lambda_{\odot}=277^{\circ}$)}

Bo\"{o}tes I (Boo I; Belokurov et al. 2006b) is located at $(l,b) = ( 358.1^{\circ}, 69.6^{\circ})$ at a distance of 62 kpc with
a radial velocity $v_{\rm GSR} = 103 \pm 3.4$ \kms (Mu{\~n}oz et al. 2006; Martin et al. 2007).
As illustrated in Figures \ref{TRAILdat.fig} and \ref{LEADdat.fig}, there is no particularly remarkable match of Boo I to any segment of the
Sgr stream, with $P_3 = 0.493$ and $P_7 = 0.774$.  
While Boo I lies in projection against the T1/T2 arms, the distance is mismatched  by 41/19 kpc respectively, and the radial velocity is mismatched
by 144/262 \kms.  Similarly, Boo I lies near the edge of the L1/L2 streams but is mismatched in distance by 11/51 kpc
and in radial velocity by 149/266 \kms. 
We therefore conclude that Boo I is not associated with the Sgr stream.

\subsection{Bo\"{o}tes II (ID=61, $\Lambda_{\odot}=277^{\circ}$)}

Bo\"{o}tes II (Boo II; Walsh et al. 2007) is located in the direction $(l,b) = (353.8^{\circ}, +68.9^{\circ})$,
and has a distance of 46 kpc and radial velocity $v_{\rm GSR} = -116$ \kms (Koch et al. 2009).
As illustrated in Figure \ref{LEADdat.fig}, 
Boo II matches the L1 arm to within $\sim 11^{\circ}$ in $B_{\odot}$,
 $\sim 70$ \kms in radial velocity, and $\sim 5$ kpc in distance.  The Sgr stream is expected to be relatively broad at this location however (i.e., near the apogalacticon of its
leading orbit), so these offsets correspond to $2\sigma, 3\sigma$, and $1\sigma$ respectively.
While Boo II is therefore moderately well matched to the Sgr stream, the statistical significance of the match is underwhelming with $P_3 = 0.124$ and $P_7 = 0.239$.
That is, in a sample of 9 ultra-faint satellites drawn from a random distribution about the Milky Way, roughly one to two such satellites (depending on whether the $P_3$ or
$P_7$ criteria are used) would be expected to match the properties
of the Sgr stream as well as Boo II by chance.
We therefore concur with Koch et al. (2009) that Boo II is perhaps the most likely of the ultra-faint dwarf satellites of the Milky Way discovered to date to be associated with Sgr,
but caution that the case for association is statistically very weak.

\subsection{Coma Berenices (ID=60, $\Lambda_{\odot}=252^{\circ}$)}

Coma Berenices (Coma Ber) was discovered by Belokurov et al. (2007) at a distance of $44 \pm 4$ kpc in the direction $(l,b) = (241.9^{\circ}, +83.6^{\circ})$,
and has a radial velocity of $v_{\rm GSR} = +81.8 \pm 0.9$ \kms (Simon \& Geha 2007).
Belokurov et al. (2009) note that Coma Ber is superimposed on the edge of the Sgr stream at a distance that suggests association with the `C' wrap
of Sgr debris in the model of Fellhauer et al. (2006).  This `C' stream corresponds to the L2 wrap of the LM10 model; as shown in Figure \ref{LEADdat.fig}, 
no such association is seen with the LM10 model with $P_3 = 0.446$ and $P_7 = 0.434$.
While Coma Ber lies in the midst of the T1/T2 wraps and at the outer edge of the L1/L2 wraps, it is mismatched
to the T1/T2/L1/L2 wraps respectively by 23/14/3/34 kpc in distance and 26/225/182/136 \kms in radial velocity.
We therefore conclude that Coma Ber is probably not associated with the Sgr system.

\subsection{Pisces I (ID=64, $\Lambda_{\odot}=73^{\circ}$)}

Pisces I was originally identified as RR Lyrae `Structure J' in the SDSS Stripe 82 catalog by Sesar et al. (2007), and recently spectroscopically confirmed
as an ultra-faint dwarf galaxy by Kollmeier et al. (2009).
Lying in the direction $(l,b) = (88^{\circ}, -58^{\circ})$, Pisces I  lies at the outer edge of the L2 debris stream,
but its large distance ($d = 85$ kpc) is inconsistent with identification with anything but the T2 wrap of the Sgr stream.  While the radial velocity of Pisces I ($v_{\rm GSR} = 45.4 \pm 4$ \kms)
is consistent with the T2 identification, its angular position is discrepant with the LM10 model of the T2 wrap by $\sim 37^{\circ}$ ($\sim 4\sigma$).
Since Pisces I therefore has formal association statistics $P_3 = 0.999$ and $P_7 = 0.290$,
we conclude that it is unlikely to be associated with Sgr.

\subsection{Segue 1 (ID=58, $\Lambda_{\odot}=225^{\circ}$)}
\label{segueI.sec}

Segue 1 was discovered in the SDSS by Belokurov et al. (2007), and lies at a distance of $23\pm2$ kpc in the direction
$(l,b) = (151.763^{\circ}, +16.074^{\circ})$.  The radial velocity of Segue 1 is $v_{\rm GSR} = 114$ \kms (Geha et al. 2009).
The nature of Segue 1 is a matter of debate: While Belokurov et al. (2007) favor the interpretation that it is an unusually extended globular cluster, the velocity dispersion
measured by Geha et al. (2009; $\sigma = 4.3 \pm 1.2$ \kms) leads these authors to conclude that Segue 1 is a dark-matter dominated dwarf galaxy.
Contradictory claims have also been made both for (Belokurov et al. 2007; Niederste-Ostholt et al. 2009) and against (Geha et al. 2009) 
the  association of Segue 1 with the Sgr tidal tails.

If Segue 1 is associated with the Sgr stream, Table \ref{results1.table} suggests that the most likely such association is with the T1 trailing wrap of debris wrapped into the northern
Galactic hemisphere (the radial velocity is mismatched with the T2/L1 arms by 190/249 \kms respectively, and the
distance is a factor of $\sim 2$ too great to match the L2 arm).
While the distance and radial velocity of Segue 1 are both relatively good matches to the T1 wrap (see Figure \ref{TRAILdat.fig}), the angular position
of Segue 1 differs from  the centroid of the stream by $18.6^{\circ}$.  If only the most recent 3 Gyr of tidal debris
(which best correspond to epochs for which there is conclusive observational evidence) are considered, we should expect the T1Y stream to be relatively narrow at this point
($\sigma_{B, {\rm Sgr}} = 4.3^{\circ}$), so that Segue 1 is offset by $\sim 4.3\sigma$ from the T1Y stream.  
If tidal debris from $> 3$ Gyr ago  is included, the angular width
of the stream increases to $\sigma_{B, {\rm Sgr}} = 8^{\circ}$, decreasing the
 discrepancy to only $\sim 2.3\sigma$.
Formally, we calculate Segue 1 to have association statistics $P_3 = 0.318$ and $P_7 = 0.113$.
Neither is particularly compelling evidence for association with the Sgr stream since $\sim 3$ such satellites (using the $P_3$ statistic) in a given sample of nine
drawn from a randomly distributed halo population are expected to match the stream better than Segue 1 simply by chance.

We note, however, that Niederste-Ostholt et al. (2009) may offer an important confirmation of the LM10 model 
of the Sgr stream at the orbital longitude of Segue 1 ($\Lambda_{\odot} = 225^{\circ}$), 
bolstering our confidence in the LM10 prediction for the location and properties of the wrapped T1 trailing arm.
In particular, their Figure 10 plots $v_{\rm GSR}$ vs. declination
for blue horizontal branch (BHB) stars from the SDSS in the range $145^{\circ} < \alpha < 155^{\circ}$ and shows two velocity peaks
at $v_{\rm GSR} = -105$ \kms and $v_{\rm GSR} = 130$ \kms.
The LM10 model suggests that these features may both be due to Sgr tidal debris, the negative velocity feature matches that expected for the L1 arm
to within 20 \kms, and the positive velocity feature matches that expected for the T1 arm to within 2 \kms. 
This agreement is particularly notable for the T1 arm since this may demonstrate that the LM10 model is consistent with
observations at least $60^{\circ}$ further along this arm than has been previously constrained by the M-giant observations of Majewski et al. (2004).
However, it is also possible that the positive-velocity feature may instead trace the Orphan stream (see, e.g., Newberg et al. 2010) which is expected to have a
similar velocity signature in this region of the sky.

Although the T1 arm may be visible in Figure 10 of Niederste-Ostholt et al. (2009), it (as indicated by the LM10 model) 
is centered around $\delta \approx 25^{\circ}$.  
Within a small field ($\lesssim 1^{\circ}$) centered on Segue 1 at $\delta=16^{\circ}$, no T1 stars should be visible and  the velocity signature of Sgr
in this small field is expected to be entirely that of the
L1 arm ($v_{\rm GSR} = -105$ \kms).
In contrast to the suggestion of Niederste-Ostholt et al. (2009), it therefore seems unlikely that recent measurements of the velocity dispersion of Segue 1
(Geha et al. 2009) have been biased by Sgr stream stars.

Some claims have also been made for the detection of tidal tails emanating from Segue 1 (Belokurov et al. 2007; although c.f. Martin et al. 2008) oriented almost
perpendicular to the Sgr stream.
If these putative tails lie along the direction of motion, their orientation may further strengthen the conclusion that Segue 1 is not associated with Sgr.  
However, at present the tails extend to such small angular distances ($\sim 2 r_{\rm eff}$; Geha et al. 2009)
that it is not obvious that they are necessarily genuine or good indicators of the orbital path of the dwarf 
(see, e.g., Johnston et al. 2002; Montuori et al. 2007; Klimentowski et al. 2009; Odenkirchen et al. 2009).
Conclusive confirmation or rejection of the hypothesis that Segue 1 is associated with the Sgr tidal stream may await the accurate determination of Segue 1's proper motion
and comparison to the values $\mu_{\alpha}$ cos$\delta = -2.4 \pm 0.4$ mas yr$^{-1}$ and $\mu_{\delta} = -1.3 \pm 0.1$ mas yr$^{-1}$ expected
for the T1 stream at this $\Lambda_{\odot}$.  At present, however, the case for Sgr association of Segue 1 does not appear to be statistically compelling.

\subsection{Segue 2 (ID=56, $\Lambda_{\odot}=119^{\circ}$)}

Segue 2 (Belokurov et al. 2009) lies 
at a distance of 35 kpc in the direction $(l,b) = (149.4^{\circ}, -38.1^{\circ})$ with a radial velocity of $v_{\rm GSR} = 43.3$ \kms.
The angular position, distance, and radial velocity of Segue 2 are overplotted on the Sgr leading and trailing tidal streams in Figures \ref{TRAILdat.fig} and \ref{LEADdat.fig}; 
we concur with Belokurov et al. (2009) in noting that Segue 2 lies near the edge of the T1 Sgr stream 
at $(\Lambda_{\odot}, B_{\odot}) = (119^{\circ}, -17^{\circ})$.  While Segue 2 also lies at roughly the same distance as the T1 stream,  
its radial velocity is inconsistent with this identification by 182 \kms ($17\sigma$).
Indeed, we calculate a $P_3 = 89.2$\% and $P_7 = 33.0$\% chance that an artificial ultra-faint satellite drawn from a random angular distribution
about the Milky Way would match the Sgr streams
better than Segue 2 by chance, leading us to strongly doubt that Segue 2 is associated with the Sgr stream.

\subsection{Segue 3 (ID=63, $\Lambda_{\odot}=41^{\circ}$)}

Segue 3 (Belokurov et al. 2010) has recently been identified as an ultra-faint star cluster at a distance of 16 kpc
in the direction $(l,b) = (69.4^{\circ}, -21.3^{\circ})$.
Although no radial velocity measurement has yet been made of this system, we estimate an association statistic $P_3 = 0.945$ based on its distance and angular position.
This extremely high value reflects the fact that
Segue 3 lies $> 30^{\circ}$ away from any segment of the Sgr stream (indeed, at $B_{\odot} = -48^{\circ}$ it does not even fall within the plot window of the uppermost panel
of Figures \ref{TRAILdat.fig} and \ref{LEADdat.fig}).
We therefore strongly doubt that Segue 3 is associated with Sgr.

\subsection{Ursa Major II (ID=57, $\Lambda_{\odot}=202^{\circ}$)}

Ursa Major II (UMa II; Zucker et al. 2006) lies at  $(l,b) = (152.5^{\circ}, +37.4^{\circ})$ with a distance of 32 kpc and a radial velocity
$v_{\rm GSR} = -33.4 \pm 1.9$ \kms (Simon \& Geha 2007).
As illustrated by Figure \ref{TRAILdat.fig}, UMa II lies 
at $(\Lambda_{\odot}, B_{\odot}) = (202^{\circ}, -33^{\circ})$ (about $3\sigma$ away from the T2 stream at $B_{\odot} = -10.9^{\circ} \pm 7.0^{\circ}$),
and is roughly consistent with the distance ($25.2 \pm 1.4$ kpc) and radial velocity ($-8.6 \pm 16$ \kms) of this segment of the stream, especially considering the large uncertainties in the
LM10 model wrapped this far from the Sgr dSph.
However, this rough consistency is not significant relative to the typical apparent correlation between artificially generated ultra-faint satellites and the Sgr streams,
with $P_3 = 0.954$ and $P_7 = 0.185$.
We are therefore inclined to doubt the association between UMa II and Sgr, although
future proper motion measurements and observational constraints on the Sgr stream at such large angular separations 
from the dSph will be required in order to conclusively
decide the matter.  If the LM10 model is still accurate for the T2 arm wrapped $\sim 360^{\circ}$ beyond the last conclusive observational constraint, we might expect the T2 stream
at the location of UMa II to have a proper motion 
$\mu_{\alpha}$ cos$\delta = -2.2 \pm 0.2$ mas yr$^{-1}$ and $\mu_{\delta} = -2.4 \pm 0.2$ mas yr$^{-1}$.

\subsection{Willman I (ID=59, $\Lambda_{\odot}=223^{\circ}$)}

Willman I (Willman et al.  2005) lies at $(l,b) = (158.57^{\circ}, +56.78^{\circ})$ with a distance of 38 kpc and a radial
velocity $v_{\rm GSR} = 35.4 \pm 2.5$ \kms (Martin et al. 2007).  As for Boo I and Coma Ber, Willman I is not obviously associated with any wrap of the Sgr stream
with $P_3 = 0.940$ and $P_7 = 0.694$.
In Figures \ref{TRAILdat.fig} and \ref{LEADdat.fig} we overplot Willman I  on the LM10 model of the Sgr stream, noting
that the satellite lies $\sim 20^{\circ}$ away from the L1/L2 wraps.
While Willman I lies nearer to the T1/T2 wraps, it is more distant than anticipated for the T1/T2 streams (28/26 kpc respectively) and has a radial velocity
inconsistent with these wraps by 102/101 \kms respectively.
We therefore conclude that Willman I is probably not associated with the Sgr system.

\subsection{Satellite Systems at Large Galactocentric Distances}
\label{distantUFs.sec}

We have omitted from consideration above the ultra-faint satellites Canes Venatici I, Canes Venatici II, Hercules, Leo IV, Leo V, Leo T, Pisces II, and Ursa Major I
since these systems lie at galactocentric distances $r_{\rm GC} > 100$ kpc at which the LM10 models predicts no tidal debris.  However, as discussed previously
in \S \ref{nearfargcs.sec}, the LM10 model does not take into account the effects of dynamical friction on the orbit of Sgr, and it is possible that stars or stellar subsystems lost
from the satellite at early times may lie at significantly larger distances than predicted by the LM10 model.  Ultra-faint dwarf satellites may also not have actually formed
in the potential well of the Sgr dwarf, but may have been gravitationally associated with it and fallen into the Milky Way as a group (e.g., Li \& Helmi 2008; D'Onghia \& Lake 2009).  
Such associated systems might be expected
to be among the first casualties of tidal stripping by the Milky Way and therefore lie at large distances relative to the present-day Sgr core
and recent tidal debris.

All of these satellites lie  out of the present orbital plane of Sgr however:
Canes Venatici I ($|Z_{\rm Sgr}| = 86$ kpc; $B_{\odot}=-23^{\circ}$),
Canes Venatici II ($|Z_{\rm Sgr}| = 53$ kpc; $B_{\odot}=20^{\circ}$), Hercules ($|Z_{\rm Sgr}| =66$ kpc, $B_{\odot}=-28^{\circ}$), Leo IV ($|Z_{\rm Sgr}| =50$ kpc, $B_{\odot}=19^{\circ}$), 
Leo V ($|Z_{\rm Sgr}| =53$ kpc, $B_{\odot}=17^{\circ}$), 
Leo T ($|Z_{\rm Sgr}| =84$ kpc, $B_{\odot}=12^{\circ}$), Pisces II ($|Z_{\rm Sgr}| =86$ kpc, $B_{\odot}=28^{\circ}$), and 
Ursa Major I ($|Z_{\rm Sgr}| =46$ kpc, $B_{\odot}=25^{\circ}$) all have $|Z_{\rm Sgr}| > 45$ kpc and $B_{\odot} > 10^{\circ}$.
Similarly, of the other Local Group dwarf galaxies only
Leo T, Leo A, Leo I, Leo II, and the Sagittarius dwarf irregular galaxy (Sgr DIG) lie within $15^{\circ}$ of the Sgr orbital plane; all of these systems lie between 200 kpc and 1 Mpc
distant from the Milky Way.
If substantial evolution has occurred in the orbit of the Sgr dSph over its lifetime, it is conceivable that some of these systems may eventually prove to be associated in some manner
with Sgr (whether forming inside the dSph or gravitationally associated with it).  However, it is not possible to meaningfully constrain this eventuality at
the present time using the LM10 model.

\section{DISCUSSION: THE SGR CONTRIBUTION TO THE GALACTIC HALO}
\label{discussion.sec}

\begin{deluxetable}{rlc}
\tablecolumns{3}
\tablewidth{0pc}
\tabletypesize{\scriptsize}
\tablecaption{Star Clusters Associated with the Sgr Stream}
\tablehead{\colhead{ID} & \colhead{Name} & \colhead{Arm} }
\startdata
\multicolumn{3}{c}{High Confidence}\\
\hline
3	 & 	Whiting 1	 & 	T1\\
27	 & 	NGC 5634	 & 	T1\\
42	 & 	M 54	 & 	CORE\\
45	 & 	Arp 2	 & 	T1\\
46	 & 	Terzan 8	 & 	T1\\
\hline
\multicolumn{3}{c}{Moderate Confidence}\\
\hline
11	& 	Berkeley 29	& L1\\
22	 & 	NGC 5053	 & 	T1\\
44	 & 	Terzan 7	 & 	T1\\
53	 & 	Pal 12	 & 	T1/L1\\
\hline
\multicolumn{3}{c}{Low Confidence}\\
\hline
7	&	Pal 2		& T1\\
18	&	NGC 4147	& T1\\
\enddata
\label{gcbudget.table}
\end{deluxetable} 

In Table \ref{gcbudget.table} we summarize the star clusters that we conclude may be associated with Sgr to high/moderate/low confidence, 
and the tidal arm in which they appear to reside.
Out of 15 candidate Sgr globular clusters selected via the criterion $P_3 < 0.15$, we find that five (Arp 2, M 54, NGC 5634, Terzan 8, and Whiting 1)
are very likely to be associated with the Sgr dwarf, an additional 4 (Berkeley 29, NGC 5053, Pal 12, Terzan 7) are moderately likely to be associated, two (NGC 4147, Pal 2)
may be associated with relatively low confidence, and four (M 53, NGC 288, Pal 5, Pal 15) are very unlikely to be associated.
Taking as our final Sgr globular cluster sample those in the high- and moderate-confidence categories, we conclude that 5-9 globular clusters are genuinely associated
with Sgr while 6-10 simply appear to be associated by chance.
This conclusion is consistent with our expectation, based on 
statistical realizations of a randomly distributed artificial globular cluster population, that $8 \pm 2$ of the 15 candidates would be genuine Sgr clusters.
As illustrated by Table \ref{prevcands.table}, the clusters which we conclude are most likely to be associated with Sgr are generally those for which there was the most consistent
agreement on Sgr membership in the literature.  We are able, however, to rule out various candidates such as M 2, NGC 288, and NGC 6356 which have also been proposed
by various authors.  Our overall {\it number} of Sgr-member globular clusters is consistent with 
Bellazzini et al. (2003a), who concluded (at 95\% confidence) that at least eight globular clusters are likely to be associated
with the Sgr system; four `outer-halo' globular clusters 
in addition to the four `core' clusters M 54, Arp 2, Terzan 7, and Terzan 8.

\begin{figure}
\plotone{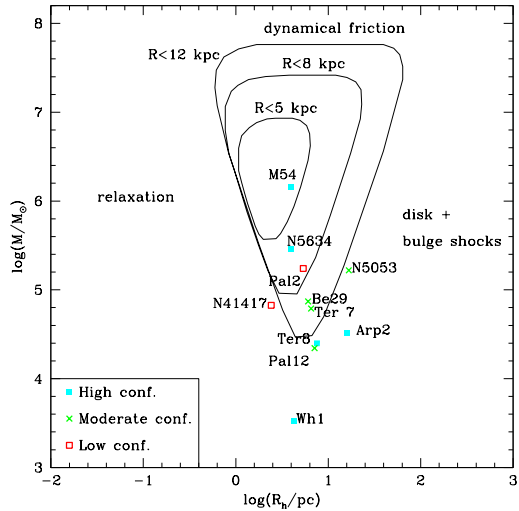}
\caption{Vital diagram (after Gnedin \& Ostriker 1997) for high, moderate, and low-confidence Sgr globular clusters.  Masses and half-light radii for the classical globular clusters 
are taken from tabulations by Gnedin \& Ostriker (1997) and Harris (1996).  These parameters are estimated for Berkeley 29 and Whiting 1 using data from Kaluzny (1994),
Tosi et al. (2004), and Carraro (2005).  Overlaid in black solid lines 
are the survival regions for distances $r_{\rm GC} =$ 12, 8, and 5 kpc from the Galactic center in the Ostriker \& Caldwell (1983) galaxy model.}
\label{survival.fig}
\end{figure}

Curiously,  of the 5-9 globular clusters which we associate
with the Sgr stream only one or two (Berkeley 29, and possibly Pal 12) 
lie in the leading arm of the Sgr stream.
In part, this may be due to selection bias; $\sim75$\% of the clusters assigned to the T1 arm lie within $\sim100^{\circ}$ of the Sgr dwarf.  The corresponding section
of the L1 arm lies almost entirely in the zone of avoidance at low Galactic latitudes where sparse clusters such as Whiting 1 and Arp 2 would be more difficult to detect.
However, this may also point to a difference in the survivability of clusters in the leading versus trailing Sgr tidal arms.
Since clusters in the leading arm would pass more closely to the Galactic center than clusters in the trailing arm, gravitational shocks from the Galactic disk may be more efficient
in disrupting clusters in the leading arm.  Indeed, as illustrated in Figure \ref{survival.fig}, sparse clusters 
such as Whiting 1 in particular
lie significantly outside the survival regime for orbits 12 kpc from the Galactic center (roughly the perigalacticon of the Sgr leading arm tidal debris on some orbits
in the triaxial halo model of LM10), and may therefore be disrupted relatively quickly once they are
removed from Sgr into the leading tidal stream.

It is perhaps unsurprising that  we are unable to associate any  stellar clusters  with the older T2/L2 wraps of Sgr tidal debris confidently, given the possible destruction of these clusters
over time, the current lack of evidence for these sections of Sgr
stream in the Galactic halo and the correspondingly poor constraints on the LM10 model in this regime.  
Our null result may, however, constitute further evidence that the Sgr dwarf has only been losing
appreciable quantities of stellar debris in the last $\sim$ 3 Gyr, which points to a constraint on the interaction time of the dwarf similar to that currently implied by wide-field observations
of the stellar debris streams.  Alternatively, discrete stellar systems such as globular clusters may simply have formed sufficiently deep in the Sgr
gravitational potential well that they have only been stripped from the dwarf after it has already experienced significant tidal disruption.

We find no compelling evidence to associate Sgr with any of the population of ultra-faint Milky Way satellites recently discovered in the SDSS.
While some of the ultra-faint satellites (e.g., Boo II, Segue 1) are roughly co-spatial with sections of the Sgr stream and have similar radial velocities, this apparent correlation
may be a product of the survey area in which the satellites were discovered; the ultra-faint satellite population is statistically consistent with 
being unassociated with the Sgr tidal streams.
It is not possible for us to discriminate between the cases in which some of these systems (a) formed within Sgr, (b) were loosely associated with Sgr and
fell into the Milky Way as a group (e.g.,  Li \& Helmi 2008; D'Onghia \& Lake 2009), or (c) are completely unrelated to Sgr.
We are therefore unable to cast any light upon the origins of these enigmatic systems, except that they were  not likely previously tightly bound parts of the Sgr dSph.

Niederste-Ostholt et al. (2010) have recently provided an estimate of the absolute magnitude of the initial Sgr dwarf by summing the light contained in the remnant Sgr 
core and the tidal streams.  Adopting the fainter of their two luminosity functions (which is most consistent with the LM10 assumption
that the present absolute magnitude of the Sgr core $M_V = -13.64$), assuming that the tidal tails are symmetric in mass content, and assuming that there is negligible luminosity
contained in regions of the stream for which there is no observational evidence in the combined 2MASS + SDSS surveys, the Niederste-Ostholt et al. (2010)
analysis indicates that the original Sgr dwarf had $M_V \approx -15.0$ (with $\sim 70$\% of this light now residing in the tidal tails).
Our 5-9 globular clusters therefore correspond to a specific globular cluster frequency
$S_N = 5 - 9$ for the original Sgr dSph, where $S_N = N_{\rm GC} \times 10^{0.4 (M_V + 15)}$ (Harris \& Van den Bergh 1981).
While this frequency is high compared to that of the Milky Way, LMC, and SMC ($S_N = 0.7, 0.8, 1.0$ respectively based on the compilation by Forbes et al. 2000), 
it is substantially less than that of Fornax
($S_N = 29$) and of previous estimates of the specific frequency of Sgr ($S_N \sim 25$; e.g., Forbes et al. 2000; Forbes \& Bridges 2010),\footnote{The main
reason for the discrepancy between our estimate of the specific frequency of Sgr and that of previous studies is our assumption of a significantly brighter
original satellite.}
and similar to that of many dE galaxies (Strader et al. 2006; Brodie \& Strader 2006, and references therein).

We caution, however, that our analysis is subject to numerous uncertainties, the greatest of which is the uncertainty in the LM10 model of the Sagittarius tidal streams.
While the LM10 model is reasonably well constrained for the most recent 3 Gyr of tidal debris for which there is the strongest evidence to date in the Galactic halo, its predictions
for the possible locations and properties of earlier epochs of tidal debris are only extrapolations.  If significant evolution has occurred in the orbit of Sgr over time
(e.g., via dynamical friction) it may be difficult to identify debris (whether stars or larger stellar systems) from the earliest epochs
whose present orbits may not resemble the current orbit of Sgr.  Our inventory of the substructures associated with Sgr may therefore be an underestimate.
In future, proper motions of the quality expected to be provided by NASA's Space Interferometry Mission (SIM) are expected to permit us to assess the case for membership
of individual stellar systems with the Sgr tidal stream with greater confidence.

\subsection{The Age-Metallicity Relation and Horizontal Branch Morphology of Sgr Globular Clusters}

\begin{figure*}
\plotone{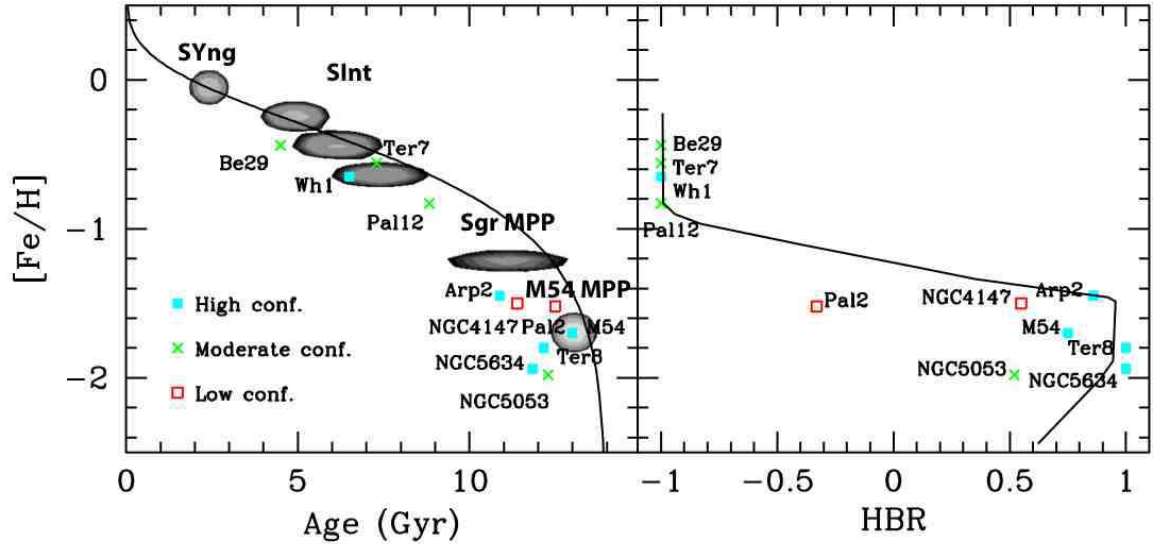}
\caption{Left panel: Age-metallicity relation for candidate Sgr globular clusters (where observational data is available).  
Filled cyan squares represent high confidence candidates, green crosses moderate confidence candidates, and red open boxes low confidence candidates.  
Overplotted is the AMR of Siegel et al. (2007; solid line), which best fits their reconstructed star formation history of Sgr (elliptical shaded regions).
Right panel: Horizontal branch morphology for candidate Sgr globular clusters as a function of metallicity.  Overlaid (solid line) is the `first-parameter' locus defined by Zinn (1993a),
objects to the left of this line are those with a significant second-parameter effect shaping their horizontal branch morphologies.}
\label{amr.fig}
\end{figure*}

In Figure \ref{amr.fig} (left panel), we plot the ages and metallicities (drawn from tabulations by Marin-Franch et al. 2009, and Forbes \& Bridges 2010)
of our eleven high/moderate/low confidence Sgr globular clusters.
As discussed recently by Forbes \& Bridges (2010) for a sample of Sgr-candidate clusters drawn from the literature, the clusters describe a relatively well defined
age-metallicity relation (AMR), which we note is
generally consistent with the closed-box Sgr AMR derived by Siegel et al. (2007).
We identify three groupings of clusters in particular in Figure \ref{amr.fig}: 
M 54\footnote{Note that the age and metallicity of M 54 was adopted from the Siegel et al. (2007) M 54 MPP measurement.}, 
NGC 5053, NGC 5634, Pal 2, and Ter 8 resemble the oldest Sgr stellar population (`M54 MPP'),
Arp 2 and NGC 4147 resemble the second-oldest Sgr stellar population (`Sgr MPP'), and Be 29, Pal 12, Ter 7, and Whiting 1
resemble the intermediate age Sgr stellar population (`SInt').

In Figure \ref{amr.fig} (right panel) we show the Lee diagram of the candidate Sgr globular clusters, plotting metallicity versus horizontal branch (HB) morphology
$HBR = (B-R)/(B+V+R)$, where $B$, $V$, and $R$ represent the number of blue, variable, and red horizontal branch stars respectively.
As has often been noted previously (e.g., Mottini \& Wallerstein 2008, and references therein), the Sgr candidate globular clusters are a bimodal group
of red- and blue-HB clusters.  
Intriguingly, the strongest Sgr cluster candidates (i.e., cyan filled squares) are those which, despite their $\sim 6$ Gyr spread in ages, 
show the least evidence for a second-parameter affecting their HB morphologies and lie closest to the single-parameter (i.e., metallicity)
locus defined by Zinn (1993a; solid line in the right panel of Figure \ref{amr.fig}).\footnote{While M 54 lies to the left
of the single-parameter locus, this is likely due to confusion between the blue HB  of M 54 and the red HB of
the more metal-rich Sgr stellar populations.}
The moderate-confidence clusters Be 29, Pal 12, and Ter 7 also follow the single-parameter locus, while NGC 5053 and the two low-confidence clusters
NGC 4147 and Pal 2 have HBs redder than expected for their metallicity.
The absence of a second-parameter effect in the Sgr globular clusters is in marked contrast to the globular cluster systems of the LMC and Fornax
(e.g., Zinn 1993b; Buonanno et al. 1999; Mackey \& Gilmore 2004), which show moderate/strong influences of a second parameter on their respective HB types at a given metallicity.
This may suggest that the strength of the second-parameter effect on the HB  morphologies of accreted globular clusters could somehow be determined by
the subhalo in which they were formed, offering a criterion by which to associate clusters in the halo of the Milky Way with their parent satellites.

Although there is a strong gradient in the chemistry of stars along the Sgr streams (see, e.g., Chou et al. 2007, 2010), there is no comparable gradient for the Sgr globular clusters.
While the two globular clusters wrapped farthest from the Sgr dwarf (NGC 5053 and NGC 5634, wrapped $266^{\circ}$ and $293^{\circ}$ from 
Sgr respectively) are the most metal-poor of the clusters, their closest chemical kin (Arp 2, M 54, Ter 8) lie $< 10^{\circ}$ from Sgr.
Similarly, the  relatively metal-rich clusters Whiting 1  and Berkeley 29 lie in the dynamically young  ($103^{\circ}$ separation from Sgr)
and old ($183^{\circ}$ separation from Sgr) sections of the stream respectively.
Indeed, if Berkeley 29 formed 4.5 Gyr ago (Carraro et al. 2004; 3.4 - 3.7 Gyr ago from the estimate of Tosi et al. 2004) 
in the `SInt' burst of star formation, the LM10 model suggests that it must have been stripped from
Sgr less than a Gyr after its formation since it would have taken $\sim 3-4$ Gyr since stripping to reach its present longitudinal separation from Sgr.

\section{SUMMARY}
\label{summary.sec}

Our major conclusions may be summarized as follows:

\begin{enumerate}
\item We find that five globular clusters (Arp 2, M 54, NGC 5634, Terzan 8, and Whiting 1)
are very likely to be associated with the Sgr dwarf, an additional 4 (Berkeley 29, NGC 5053, Pal 12, Terzan 7) are moderately likely to be associated, and two (NGC 4147, Pal 2)
may be associated with relatively low confidence.
Confining our attention to  those in the high- and moderate-confidence categories, we conclude that 5-9 globular clusters are genuinely associated
with Sgr, consistent with our expectation of $8\pm2$ clusters based on
statistical realizations of a randomly distributed artificial globular cluster population.

\item Based on updated estimates of the initial luminosity of the pre-interaction Sgr dwarf ($M_V = -15$), we estimate a specific globular cluster frequency $S_N = 5 - 9$
typical of dE galaxies.

\item The globular clusters identified as most likely to be associated with Sgr are consistent with the AMR of Sgr's stellar populations, and show no evidence for a second-parameter
effect shaping their horizontal branch morphologies.  If the presence of a second-parameter effect is a reliable indicator of Sgr membership, it suggests that the two lowest-confidence
Sgr clusters (NGC 4147 and Pal 2) and the moderate-confidence cluster NGC 5053 may not be genuinely associated with Sgr.  There is no obvious correlation
of age, metallicity, or horizontal branch morphology with distance from Sgr along the tidal streams.

\item We find no compelling evidence to associate Sgr with any of the population of ultra-faint Milky Way satellites recently discovered in the SDSS.
The ultra-faint satellite population is statistically consistent with being unassociated with the Sgr tidal streams, but we are unable to rule out association conclusively in all cases.

\end{enumerate}

We caution, however, that our results are based upon comparisons to sections of the Sgr stream that we presently observe in the Galactic halo, and do not account for possible
variations in the orbit of Sgr over time.  For instance, we do not address here the (likely) possibility that dynamical friction has altered the orbit of Sgr over its lifetime, nor are we able to
constrain the possibility that some objects may have been loosely affiliated with Sgr and fallen into the Milky Way as a group.  Future observational constraints on the length and
characteristics of the Sgr tidal streams will help to further understand the accretion origins of substructure within our Galactic halo.

\acknowledgements

Support for this work was provided by NASA through Hubble Fellowship grant \# HF-51244.01
awarded by the Space Telescope Science Institute, which is operated by the Association of Universities for Research in Astronomy, Inc., for NASA, under contract NAS 5-26555.

\end{document}